\documentstyle[12pt]{article}
\makeatletter \oddsidemargin  -.1in \evensidemargin -.1in
\newcommand{\singlespacing}{\let\CS=\@currsize\renewcommand{\baselinestretch}{1}\tiny\CS}
\newcommand{\oneandahalfspacing}{\let\CS=\@currsize\renewcommand{\baselinestretch}{1.25}\tiny\CS}
\newcommand{\doublespacing}{\let\CS=\@currsize\renewcommand{\baselinestretch}{1.35}\tiny\CS}

\newtheorem{rule-def}[theorem]{Rule}

\RequirePackage[dvips]{graphicx} \textheight 21.5cm

\begin{document}

\title{\bf Combined effects of thermal radiation and Hall current on MHD
free-convective flow and mass transfer over a stretching sheet
with variable viscosity}
\author{\small G. C. Shit\thanks{Email address: gcs@math.jdvu.ac.in
(G. C. Shit)}~~ and ~~R. Haldar \\
\it Department of Mathematics\\
\it Jadavpur University, Kolkata - 700032, India\\
}
\date{}
\maketitle \noindent \doublespacing
\begin{abstract} An analysis has been investigated for the effects of thermal radiation
and Hall current on magnetohydrodynamic free-convective flow and
mass transfer over a stretching sheet with variable viscosity in
the presence of heat generation/absorption. The fluid viscosity is
assumed to vary as an inverse linear function of temperature. The
boundary-layer equations governing the flow problem under
consideration have been reduced to a system of non-linear ordinary
differential equations by employing a similarity transformation.
Using the finite difference scheme, numerical solutions to the
transform ordinary differential equations have been solved and the
results that obtained are presented graphically. With an aim to
test the accuracy, the numerical results have been compared with
the existing scientific literature and found excellent agreement.  \\

\end{abstract}

\section {Introduction}
In the recent past there has been a growing interest in
boundary-layer flow on a continuous moving surface in the presence
of magnetic field with or without considering the effect of
 Hall current. But, of course these are very significant type of
flow towards the several engineering applications, such as in
polymer processing, electro-chemistry, MHD power generators,
flight magnetohydrodynamics as well as in the field of planetary
magnetosphere, aeronautics and chemical engineering. \\

Sakiadis \cite{Sakiadis} first explored  the study of
boundary-layer flow on a continuous moving surface and Crane
\cite{Crane} extended this problem to a stretching sheet whose
surface velocity varies linearly with the distance $x$ from a
fixed origin. During the past decades  several investigators
[3-10] have considered the boundary-layer flow problems under
different physical situations. Gupta and Gupta \cite{Gupta and
Gupta} examined the heat and mass transfer over a stretching sheet
subject to suction or blowing. The influence of a uniform magnetic
field on the flow of an electrically conducting fluid past a
stretching sheet was investigated by Pavlov \cite{Pavlov},
Andersson \cite{Andersson2}, Andersson et al.\cite{Andersson1},
Gupta and Chakrabarty \cite{Gupta and Chakrabarty}, Char
\cite{Char}, Watanabe and Pop \cite{Watanabe and Pop1} and
Elbashbeshy \cite{Elbashbeshy}. The chemical reaction on
free-convective flow and mass transfer of a viscous,
incompressible and electrically conducting fluid over a stretching
sheet was investigated by Afify \cite{Afify} in the presence of a
uniform transfer magnetic field. In all these investigations the
electrical conductivity of the fluid was assumed to be uniform and
low magnetic field intensity. However, in an ionized fluid where
the density is low and thereby magnetic field intensity is very
strong, the conductivity normal to the magnetic field is reduced
due to the spiraling of electrons and ions about the magnetic
lines of force before collisions take place and a current induced
in a direction normal to both the electric and magnetic fields.
This phenomena is known as Hall effect. Watanabe and Pop
\cite{Watanabe and Pop} investigated the magnetohydrodynamic
boundary-layer flow over a continuously moving semi-infinite flat
plate by taking into account the Hall currents. Aboeldahab
\cite{Aboeldahab} and Aboeldahab and Elbarbary \cite{Aboeldahab
and Elbarbary} studied the Hall current effects on MHD
free-convective flow past a vertical plate with mass transfer.
Shit \cite{Shit} investigated the Hall effects on MHD
free-convective flow and mass transfer over a stretching sheet in
the presence of chemical reaction. Fakhar et al. \cite{Fakhar}
studied the Hall effects on the unsteady magnetohydrodynamic flow
of a third grade fluid without considering the heat and mass
transfer phenomena. The effect of Hall currents on the steady MHD
flow of Berger's fluid between two parallel electrically
insulating infinite planes was carried out by Rana et al.
\cite{Rana}.\\

Recently, a new idea is added to the study of boundary-layer fluid
flow and heat transfer is the consideration of the effect of
thermal radiation and temperature dependent viscosity. Many
processes in engineering applications occur at high temperatures
and the radiate heat transfer becomes very important for the
design of the pertinent equipment. In view of this, Rapits and
Perdikis \cite{Rapits and perdikis} and Rapits \cite{Rapits}
studied respectively the flow of a visco-elastic fluid and
micropolar fluid past a stretching sheet in the presence of
thermal radiation. Mukhopadhaya et al. \cite{Mukhopadhaya}
investigated the problem of MHD boundary-layer flow over a heated
stretching sheet with variable viscosity. The radiation effect on
boundary-layer flow with or without applying magnetic field under
different situations were studied by Shateyi \cite{ Shateyi},
Mahmoud \cite{Mahmoud}, Pal and Talukdar \cite{Pal and Talukdar}
and Pal and Chatterjee \cite{Pal and Chatterjee}. However, Salem
\cite{Salem} investigated the effect of variable viscosity on MHD
viscoelastic fluid flow and heat transfer over a stretching sheet
without considering thermal radiation effect. Shit and Haldar
\cite{Shit and Haldar} carried out the study of the effect of
thermal radiation on MHD viscoelastic fluid flow past a stretching
surface with variable viscosity. But, no attempt is available in
the existing scientific literatures for the consideration of the
combined effect of thermal radiation and Hall current on the study
of MHD boundary-layer flow. Thus the present study fills the gap
in this directions.\\

 Again, combined heat and mass transfer
problems with chemical reaction are of increasing importance in
many processes, like drying, evaporation at the surface of a water
body, energy transfer in a wet cooling tower etc. In this context,
Muthucumarswamy and Ganesan \cite{Muthucumarswamy and Ganesan}
studied the effect of the chemical reaction and injection as well
as flow characteristics in an unsteady upward motion of an
isothermal plate. Chamakha \cite{Chamakha} carried out the MHD
flow of uniformly stretching vertical permeable surface in the
presence of heat generation / absorption along with the chemical
reaction. Very recently, Mohamed and Abo-Dahab \cite{Mohamed and
Abo-Dahab} have investigated the influence of chemical reaction
and thermal radiation on hydromagnetic free-convective heat and
mass transfer for a micropolar fluid via a porous medium bounded
by an infinite vertical porous plate in the presence of heat
generation. Seddeek et al. \cite{Seddeek1, Seddeek2} analyzed the
effects of chemical reaction, radiation and variable viscosity on
hydromagnetic mixed convection heat and mass transfer. In all
these studies ignores the effect of
the consideration of Hall current.\\

Now we propose to study the combined effects of thermal radiation
and Hall current on the hydromagnetic free-convective flow and
mass transfer over a stretching surface with variable viscosity in
the presence of heat generation/absorption. The present problem
pertains to situation in which the n-th order chemical reaction
takes place. Thus the study is also applicable to the elongation
of the bubbles and in bioengineering where the flexible surfaces
of the biological cells and membranes in living systems are
typically surrounded with fluids which are electrically conducting
and being stretched constantly.

\section{Mathematical Formulation}
We consider the steady free-convective flow and mass transfer of
an incompressible, viscous and electrically conducting fluid past
a stretching sheet and the sheet is stretched with a velocity
proportional to the distance from a fixed origin O (cf. Fig. 1). A
uniform strong magnetic field of strength $ B_0$ is imposed along
the $y$-axis and the effect of Hall currents is taken into
account.  Taking Hall effects into account the generalized Ohm's
law \cite{Cowling} may
be put in the form :\\
   \[ \overrightarrow{J} =\frac{\sigma}{1+m^2} \left(\overrightarrow{E}+
   \overrightarrow{V}\times \overrightarrow{B} -\frac{1}{e n_e} \overrightarrow{J}\times
   \overrightarrow{B}\right),  \] \\
 where $\overrightarrow{V}$ represents the velocity vector, $\overrightarrow{E}$ is the intensity
 vector of the electric field, $\overrightarrow{B}$ is the magnetic induction
 vector, $\nu_e$ the magnetic permeability, $\overrightarrow{J}$ the electric
 current density vector, $m=\frac{\sigma B_0}{e n_e}$ is the Hall
 current parameter, $\sigma$ the electrical conductivity, $e$ the charge
 of the electron and $n_e$ is the number density of the electron. A very interesting is that effect
 of Hall current gives rise to a force in the $z$-direction which in turn produces a
 cross flow velocity in this direction and thus the flow becomes
 three-dimensional. \\
 The temperature and the species concentration are
maintained at a prescribed constant values $T_w$, $C_w$ at the
sheet and $T_\infty$ and $C_\infty$ are the fixed values far away
from the sheet. Since the concentration of diffusing species is
very small in comparison to other chemical species, the Soret and
Dufour effects
are neglected.\\
 Following Lai and Kulacki \cite{Lai and Kulacki}, the fluid
viscosity $\mu$ is assumed to vary as a reciprocal of a linear
function of temperature given by

 \begin{eqnarray}
 \frac{1}{\mu}
 =\frac{1}{\mu_{\infty}}\left[1+\gamma(T-T_{\infty})\right ]
 \end{eqnarray}
 or
 \begin{eqnarray}
 \frac{1}{\mu_{\infty}}=a({T-T_r})
 \end{eqnarray}
 \noindent where
      $ a=\frac{\gamma}{\mu_{\infty}}
  ~~~and~~~    T_r={T_\infty-\frac{1}{\gamma}}$.

\noindent In the above equation both $a$ and $T_r$ are constants,
and their values depend on the thermal property of the fluid,
i.e., $\gamma$. In
general $a> 0$ represent for liquids, whereas for gases $a< 0$.\\
    By assuming Rosseland approximation for radiation, the
    radiative heat flux $q_r$ is given by
    \begin{eqnarray}
    q_r = -\frac{4 \sigma^*}{3 K}\frac{\partial T^4}{\partial y}
    \end{eqnarray}
    \noindent  where $\sigma^*$ is the Stefan-Boltzman constant and $K$ the
    mean absorption coefficient. We assume that the temperature
    differences within the flow are sufficiently small such that
    $T^4$ may be expressed as a linear function of the temperature
    as shown in Chamakha \cite{Chamakha}.
    Expanding $T^4$ in a Taylor series about $T_\infty$ and
    neglecting higher order terms, we obtain
    \begin{eqnarray}
    T^4 \cong 4 T^3_\infty T-3 T^4_\infty
    \end{eqnarray}
    Substituting $T^4$ from (4) in (3) and differentiating the resulting equation
     with respect to $y$, we obtain as
\begin{eqnarray}
   \frac{\partial q_r}{\partial y} = - \frac{16 \sigma^*
   T^3_\infty}{3 K}\frac{\partial^2 T}{\partial y^2}
    \end{eqnarray}
 Owing to the above mentioned assumptions, the boundary layer
 free-convection flow with mass transfer and generalized Ohm's law
 is governed by the following system of equations :
\begin{eqnarray}
\frac{\partial u}{\partial x} + \frac{\partial v}{\partial y}
=0,~~
\end{eqnarray}

\begin{eqnarray}
{\rho_{\infty}}\left(u\frac{\partial u}{\partial
x}+v\frac{\partial u}{\partial y}\right)=\frac{\partial }{\partial
y}\left ( \mu \frac{\partial u}{\partial y}\right )
+{\rho_{\infty}}g\beta_t \left ( T-T_\infty \right )
&+&{\rho_{\infty}}g\beta_c \left (C-C_\infty \right )\nonumber\\
&-&\frac{\sigma B_0^2}{1+m^2}\left(u+mw \right),
\end{eqnarray}

\begin{eqnarray}
{\rho_{\infty}}\left(u\frac{\partial w}{\partial
x}+v\frac{\partial w}{\partial y}\right)=\frac{\partial }{\partial
y}\left ( \mu \frac{\partial w}{\partial y}\right )+\frac{\sigma
B_0^2}{1+m^2}\left(mu-w \right),
\end{eqnarray}

\begin{eqnarray}
{\rho_\infty c_p}\left (u\frac{\partial T}{\partial
x}+v\frac{\partial T}{\partial y}\right )=k\frac{\partial ^2
T}{\partial y^2}-Q \left (T-T_\infty \right )-\frac{\partial
q_r}{\partial y},
\end{eqnarray}

\begin{eqnarray}
u\frac{\partial C}{\partial x}+v\frac{\partial C}{\partial y}=D
\frac{\partial ^2 C}{\partial y^2}-k_0\left ( C-C_\infty \right
)^{n},
\end{eqnarray}
where $(u, v, w)$ are the velocity components along the $(x, y, z)
$ directions respectively, $\mu $ is the coefficient of viscosity,
$g $ the acceleration due to gravity, $\beta_t $ the coefficient
of thermal expansion, $\beta_c $ the coefficient of expansion with
concentration, $T $ and $C $ are the temperature and concentration
respectively, $D$ the thermal molecular diffusivity, $k_0$ is the
constant measures the rate of reaction, $\sigma$ is the electrical
conductivity, $c_p$ is the specific heat at constant pressure, $k
$ is the thermal conductivity, $T_\infty $ and $\rho_\infty $ are
the free stream temperature and density and
$n$ be the order of reaction. \\

\noindent The boundary conditions to the present problem can be
written as
\begin{eqnarray}
u=bx,~~ v=w=0,~~ T=T_w,~~ C=C_w~~ at ~~y=0
\end{eqnarray}
\begin{eqnarray}
u \rightarrow 0,~~ w \rightarrow 0,~~T \rightarrow T_\infty, ~~ C
\rightarrow C_\infty ~~as ~~ y \rightarrow \infty
\end{eqnarray}
where $b(>0)$ being stretching rate of the sheet. The boundary
conditions on velocity in (11) are the no-slip condition at the
surface $y=0$, while the boundary conditions on velocity at $y
\rightarrow 0$ follow from the fact that there is no flow far way
from the stretching surface. \\

\noindent To examine the flow regime adjacent to the sheet, the
following transformations are invoked
\begin{eqnarray}
u=bxf'(\eta); ~~ v=- \sqrt{b\nu}f(\eta); ~~ w=bxg(\eta); ~~ \eta
=\sqrt{\frac{b}{\nu}}y; ~~\theta(\eta)
=\frac{T-T_\infty}{T_w-T_\infty};~~\phi(\eta)
=\frac{C-C_\infty}{C_w-C_\infty}
\end{eqnarray}

 \noindent where $f$ is a dimensionless stream function, $\eta$ is the
similarity space variable, $\theta$ and $\phi$ are the
dimensionless temperature and concentration respectively. Clearly,
the continuity equation (6) is satisfied by
 $u$ and $v $ defined in (13) on substitution which
 into equations (7) - (10) gives

\begin{eqnarray}
 \left (\frac{\theta-\theta_r}{\theta_r}\right )\left (f'^2-ff''\right ) +f'''
 - \left (\frac{\theta'}{\theta-\theta_r}\right )f''
 &-&\left (\frac{\theta-\theta_r}{\theta_r}\right )(Gr  \theta +Gc
 \phi)\nonumber\\&+&M\left (\frac{\theta-\theta_r}{\theta_r}\right )\left (\frac{ f'+m
 g}{1+m^2}\right)=0,
\end{eqnarray}

\begin{eqnarray}
 \left (\frac{\theta-\theta_r}{\theta_r}\right )\left (f'g-fg'\right
 )+g'' - \left (\frac{\theta '}{\theta-\theta_r}\right) g'-
M\left (\frac{\theta-\theta_r}{\theta_r}\right )\left (\frac{m f'-
g}{1+m^2}\right) =0,
\end{eqnarray}

\begin{eqnarray}
 (3Nr +4)\theta'' +3NrPrf\theta' +3NrPr\lambda \theta =0,
 \end{eqnarray}

 \begin{eqnarray}
  \phi'' +Sc\left(f\phi'-\gamma\phi^n\right ) =0,
  \end{eqnarray}

 \noindent and the transformed boundary conditions are given by
  \begin{eqnarray}
    f'(\eta) =1,~~ f(\eta) =0,~~ g(\eta) =0,~~  \theta(\eta) =1,~~ \phi(\eta) =1~~ at
    ~~  \eta =0,
    \end{eqnarray}
    \begin{eqnarray}
    f'(\eta) =0, ~~g(\eta) =0 , ~~\theta(\eta) =0, ~~\phi(\eta) =0 ~~ at
   ~~ \eta \rightarrow \infty,
 \end{eqnarray}

  \noindent where primes denotes differentiation with respect to
  $\eta $ only and the dimensionless parameters appearing in the
  equations (14)-(17) are respectively
  $ \theta_r=\frac{T_r-T_\infty}{T_w-T_\infty}
  = -\left[\frac{1}{\gamma(T_w-T_\infty)}\right] $  is known as the viscosity parameter,
   $M = \frac{\sigma B_0^2}{\rho_\infty b} $  the magnetic
  parameter, $P_r = \frac {\rho_\infty C_p \nu}{k} $ the Prandtl
  number, $m=\frac{\sigma B_0}{e n_e} $ is the Hall current
  parameter,$\gamma=\frac{k_0}{b} \left (C_w-C_\infty \right)^{n-1}$
  the non-dimensional chemical reaction parameter,
  $G_r = \frac {{\rho_\infty }g \beta_t (T_w-T_\infty)}{b^2 x} $ the Grashof
  number, $G_c = \frac{{\rho_\infty }g \beta_c(C_w-C_\infty)}{b^2 x} $  the
  modified Grashof number, $ Nr = \frac{k K}{4 T^3_\infty \sigma^*} $
  the thermal radiation parameter,
   $Sc = \frac{\mu}{\rho_\infty D} $ the Schmidt number and $\lambda = \frac{Q}{\rho_\infty C_p b}$
   is defined as the heat generation/absorption parameter. \\

 \noindent The important characteristics of the present investigation are the
   local skin-friction coefficient $C_f$, the local Nusselt
  number  $ Nu $ and the local Sherwood number $Sh$ defined by
   \begin{eqnarray}
     C_f = \frac{\tau_w }{\mu bx\sqrt{\frac{b}{\nu}}}=
      f''(0),
      ~where~~\tau_w = {\mu}\left(\frac{\partial u}{\partial y}\right )_{y=0}
      ={\mu bx\sqrt{\frac{b}{\nu}}}{ f''(0)},~~
   \end{eqnarray}
   \begin{eqnarray}
   Nu=\frac{q_w}{k\sqrt{\frac{b}{\nu}}(T_w-T_\infty)}=-\theta'(0), ~~where~~ q_w=-k\left
(\frac{\partial T}{\partial y} \right
)_{y=0}=-k\sqrt{\frac{b}{\nu}}(T_w-T_\infty)\theta'(0),~~~
\end{eqnarray}
\begin{eqnarray}
Sh=\frac{m_w}{D\sqrt{\frac{b}{\nu}}(C_w-C_\infty)}=-\phi'(0),
~~where~~ m_w=-D\left (\frac{\partial C}{\partial y} \right
)_{y=0}=-D\sqrt{\frac{b}{\nu}}(C_w-C_\infty)\phi'(0),~~~
\end{eqnarray}\\
If we consider $M=m=0$ and $Nr=Sc=G_r=G_c=0$, the present flow
problem becomes hydrodynamics boundary-layer flow past a
stretching sheet whose
analytical solution put forwarded by Crane \cite{Crane} as follows :\\
\begin{eqnarray}
f(\eta)=1-e^{-\eta}~~~ i.e,~~~f'(\eta)=e^{-\eta}
\end{eqnarray}\\
An attempt has been made to validate our results for the axial
velocity $f'(\eta)$, we compared our results with this analytical
solution and have found excellent agreement.\\

\section{Numerical Results and Discussion}
The system of coupled and non-linear ordinary differential
equations (14)-(17) along with the boundary conditions (18) and
(19) have been solved numerically by employing a finite difference
scheme. We used Newton's linearization method (cf. Cebeci and
Couteix \cite{Cebeci and Couteix}) to linearize the discretized
equations. The essential features of this technique is that it is
based on a finite difference scheme, which has better stability,
simple, accurate and more efficient. Finite difference technique
leads to a system which is tri-diagonal and therefore speedy
convergence as well as economical memory space to store the
coefficients. The computational work has been carried out by
taking $\delta\eta =0.0125$ and further reduction in $\delta\eta$
does not bring about any significant change. In the present study,
the numerical values to the physical parameters have been chosen
so that $M, m, Nr, \theta_r, Pr, Gr, Gc, Sc, n, \gamma $ and
$\lambda$ are varied over a range, which are listed in the figure
legends. Fig. 2 shows that our numerical results are
complete agreement with those of Crane \cite{Crane}.\\

Figs. 3 - 10 illustrate the variation of axial velocity for
different values of the dimensionless parameters that involved in
the present study. Fig. 3 shows that the axial velocity decreases
with the increase of the magnetic parameter $M$, whereas from Fig.
4 it indicates that the axial velocity increases with the increase
of Hall parameters $m$. This is due to the fact that as $M$
increases, the Lorentz force which opposes the flow and leads to
deceleration of the fluid motion. By contrast, the cross flow
velocity component $g(\eta) $ induced due to Hall effects and
shows a anomalous behaviour in $f'(\eta) $ with the variation of
$M$. It has been shown in Figs 5 and 6 that the axial velocity
decreases with the increase of the Prandtl number $Pr$ as well as
the thermal radiation parameter $Nr$. This is due to fact that
there would be a decrease of boundary-layer thickness in the
presence of thermal radiation. Fig. 7 depicts that the axial
velocity $f'(\eta)$ increases with the decreasing of the viscosity
parameter $\theta_r$. This observation leads to an increase of the
thermal boundary-layer thickness. The effects of heat generation
parameter $(\lambda>0)$ and absorption parameter $(\lambda<0)$ on
the axial velocity displayed in Fig. 8. This figure shows that the
axial velocity decreases as the parameter $\lambda$ increases. The
variation of Schmidt number $Sc$ and the chemical reaction
parameter $\gamma$ on the axial velocity profile $f'(\eta)$ shown
in Figs. 9 and 10 respectively. It is obvious that the increased
values of $Sc$ and $\gamma$ tend
to decreasing of the velocity profiles across the boundary-layer.\\

Figs. 11 -19 give the distribution of the $z$- component of
velocity, which is induced due to the presence of Hall effects.
All these figures show that for any particular values of the
physical parameters $g(\eta)$ reaches a maximum value at a certain
height $\eta$ above the sheet and beyond which $g(\eta)$ decreases
gradually in asymptotic nature. It is noticed from Fig. 11 is that
in the absence of magnetic parameter $M ( = 0)$ cross flow
velocity vanishes. This is due to fact that, when there is no
applied magnetic field, the cross flow velocity would not arise.
The variation of Hall current parameter $m$ on the cross-flow
velocity $g(\eta)$ shown in Fig. 12. An interesting result
observed from this figure that the cross-flow velocity gradually
increases with the increase of $m\leq 2$ and the velocity
decreases for $m> 2$. The values of $m$ beyond which the flow
behaviour changes is considerable depend upon the choice of the
magnetic parameter $M$. Thus we conclude that after certain
magnetic field strength the flow behaviour is significantly
affected. Figs. 13 and 14 indicate that the cross-flow velocity
$g(\eta)$ decreases with increasing the Prandtl number $Pr$ and
the thermal radiation parameter $Nr$, while from Fig. 15 that the
trend is reversed after a certain height above the sheet. This is
lies in the fact that the increase of Prandtl number $Pr$ and the
thermal radiation $Nr$ give rise to decrease of the momentum
boundary-layer thickness. We observed from Figs. 16, 17 and 19
that the cross-flow velocity decreases with the increase of the
heat generation/absorption parameter $\lambda$, Schmidt number
$Sc$ and the chemical reaction parameter $\gamma$. However, the
cross-flow velocity increases as the order of the chemical
reaction $n$ increases.\\

 The distribution of dimensionless
temperature $\theta(\eta)$ along the height of the stretching
sheet for different values of the dimensionless parameters
involved in the present study displayed through Figs. 20 - 25.
Fig. 20 shows that by the application of an external magnetic
enhances the temperature of the fluid, while the effect of the
Hall current parameter $m$ has an reducing effect on the
dimensionless temperature $\theta(\eta)$ shown in Fig. 21. It may
note  that the effect of Hall current parameter $m$ opposes the
effect of magnetic field on the temperature distribution. Fig. 22
presents the variation of Prandtl number $Pr$ on the temperature
$\theta(\eta)$. The results presented in Fig. 22 shows that the
dimensionless temperature decreases as the Prandatl number $Pr$
increases. This is lies in the fact that smaller values of $Pr$
are equivalent to increasing of thermal conductivities and
therefore heat is able diffuse away from the stretching sheet.
Fig. 23 represents the temperature profiles for various values of
the thermal radiation parameter $Nr$ in the boundary-layer.
Increasing the thermal radiation parameter $Nr$ produces a
decrease in the temperature of the fluid. This is because of the
fact that the thermal boundary-layer thickness decreases with
increasing  the thermal radiation parameter. Fig. 24 gives the
variation  of the viscosity parameter $\theta_r$ on the
temperature profiles and which shows that  no significant change
occur with the increase of the values of $\theta_r$. It is evident
that the energy equation (16) is uncoupled from the viscosity
parameter $\theta_r$. For this reason no figures are presented
herein for the variation of $Sc$, $n$ and $\gamma$. The effect of
heat generation/absorption parameter $\lambda$ on the
dimensionless temperature $\theta(\eta)$ is shown in Fig. 25. It
is clear that an increase in the heat generation/absorption
parameter $\lambda$ leads to a decrease of $\theta(\eta)$. Thus
the effect of internal heat generation is to decrease the rate of
energy transport to the fluid, thereby decreasing the temperature
of the fluid.\\

The effect of the imposition of various parameters on
concentration profiles are shown in Figs. 26 - 31. It is observed
from Figs. 26 and 27 that under the action of a strong magnetic
field, the concentration species has an enhancing effect, whereas
it has the reducing effect on the Hall current parameter $m$. The
reason behind this is that in equations (14) and (15), the
parameters $M$ and $m$ are connected by the relations of the form
$\frac{M}{1+m^2}$ and $\frac{M m}{1+m^2}$. As the variation of
concentration profiles for different values of the parameters $Nr$
and $\theta_r$ agrees with the temperature profiles, we do not
have presented those results. However, Fig. 28 is the variation of
heat generation/absorption parameter $\lambda$ on the species
concentration, which opposes  the variation of the temperature
distribution. Fig. 29 depicts the variation of Schmidt number $Sc$
on the species concentration $\phi(\eta)$. It is observed that the
species concentration decreases with the increase of the Schmidt
number $Sc$. Physically, which shows that the increase of $Sc$
causes decrease of molecular diffusion $D$. The influence of
chemical reaction parameter $\gamma$ on the species concentration
profiles for generative chemical reaction is shown in Fig. 30. It
is noticed that the increasing values of the chemical reaction
parameter $\gamma$ is to decelerate the concentration distribution
in the boundary-layer. This is due to fact that destructive
$(\gamma>0)$ chemical reaction decreases the shortest
boundary-layer thickness and thereby increasing the species
concentration. It is observed from Fig. 31 that the increasing
effect of the order of chemical reaction parameter  $n$ is to
enhances the mass transfer in the boundary-layer.\\

 The problems of engineering interest for the present study are the local
skin-friction coefficient $C_f$ , the local Nusselt number $Nu$
and the local sherwood number $Sh$ which indicates physically the
wall shear stress, the rate of heat transfer at the sheet and the
rate of mass transfer respectively. The expressions of these
physical quantities have been presented in equations (20)- (22).
Tables 1-3 exhibit the numerical values to the local skin-friction
coefficient $f''(0)$, the local Nusselt number $-\theta'(0)$ and
the local sherwood number $-\phi'(0)$ respectively. It has been
observed that for any particular values of the parameters $Pr$,
$Nr$, $m$, $\theta_r$, $\gamma$ , $\lambda$ and $Sc$, the local
skin-friction coefficient, the local Nusselt number and the local
Shearwood number decreases with the increase of the magnetic
parameter $M$. Table -1 shows that for fixed value of $M$, the
local skin-friction $C_f$ decreases with increasing the values of
the Prandatl number $Pr$, thermal radiation parameter $Nr$,
viscosity parameter $\theta_r$, the rate of chemical reaction
$\gamma$, heat generation/absorption parameter $\lambda$ and the
Schmidt number $Sc$. Physically, we meant that for the increasing
of the thermal radiation parameter $Nr$ leads to decreasing in the
boundary-layer thickness. However, the skin-friction coefficient
increases as the Hall parameter $m$ increases. It is worthwhile to
observed from Table-2 that the rate of heat transfer at the sheet
increases with the increase of the parameters $Pr$, $Nr$, $m$ and
$\gamma$. The increase of the thermal radiation parameter $Nr$ has
an enhancing effect on the thermal boundary-layer thickness. It is
also noticed that the increasing of the parameters $\theta_r$,
$\gamma$ and $Sc$ is to reduce the heat transfer rate and which
shows that the rate of change of heat transfer is insignificant.
The rate of mass transfer increases significantly when the Hall
parameter $m$, the chemical reaction parameter $\gamma$ and the
Schmidt number $Sc$ increases. Moreover, the rate of mass transfer
decreases insignificantly with
the increase of the parameters $Pr$, $Nr$ and $\lambda$. \\

\section{Conclusions}
In the present investigation, we dealt with the combined effects
of thermal radiation and Hall current on the boundary-layer fluid
flow, heat and mass transfer with temperature dependent viscosity.
The highly non-linear coupled system of partial differential
equations characterizing the flow, heat and mass transfer has been
reduced to a coupled system of non-linear ordinary differential
equations by applying a suitable similarity transformations. The
resulting system solved numerically by using the finite difference
scheme along with the Newton's linearzation technique. The
obtained numerical results have been presented through the figures
and in tabular form to illustrate the details of the flow
behaviour, heat and mass transfer phenomena and their dependence
on the physical parameters that involved in the present
investigation. From our computed numerical results we observed
that the magnetic field and Hall current produce opposite effects
on the velocity distribution and heat transfer as well as on the
concentration distribution. For a fixed value of $M$, the
skin-friction increases with an increase in $m$ and similar is the
observation for heat and mass transfer rate. Sufficiently strong
heat generation parameter may alter the temperature gradient.
Temperature decreases and the concentration increases with
increasing values of Prandtl number $Pr$. But a reversal trend is
observed when the values of the thermal radiation parameter
increases. The species concentration decreases with an increase in
the values of the Schmidt number $Sc$ and chemical reaction
parameter $\lambda$ whereas opposite trend is observed in the case
of the order of the chemical reaction. It is hopped that the
results obtained will serve as a scientific tool for understanding
more complex flow problems and provide more useful
 information for engineering applications.\\

\newpage
\textheight 22.0cm \pagebreak
\begin{minipage}{1.0\textwidth}
   \begin{center}
      \includegraphics[width=3.8in,height=2.5in ]{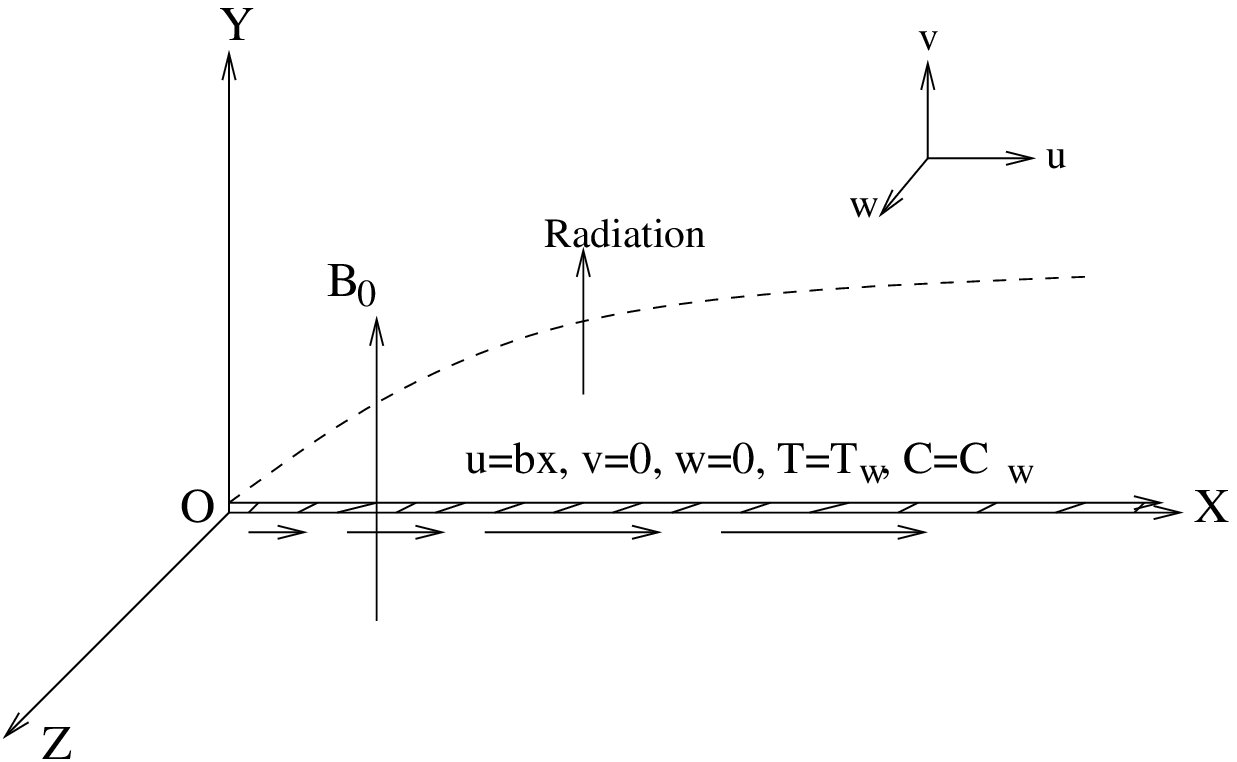}\\
Fig. 1 ~~ Physical sketch of the problem \vspace*{1.5cm} \\

      \includegraphics[width=3.8in,height=2.5in ]{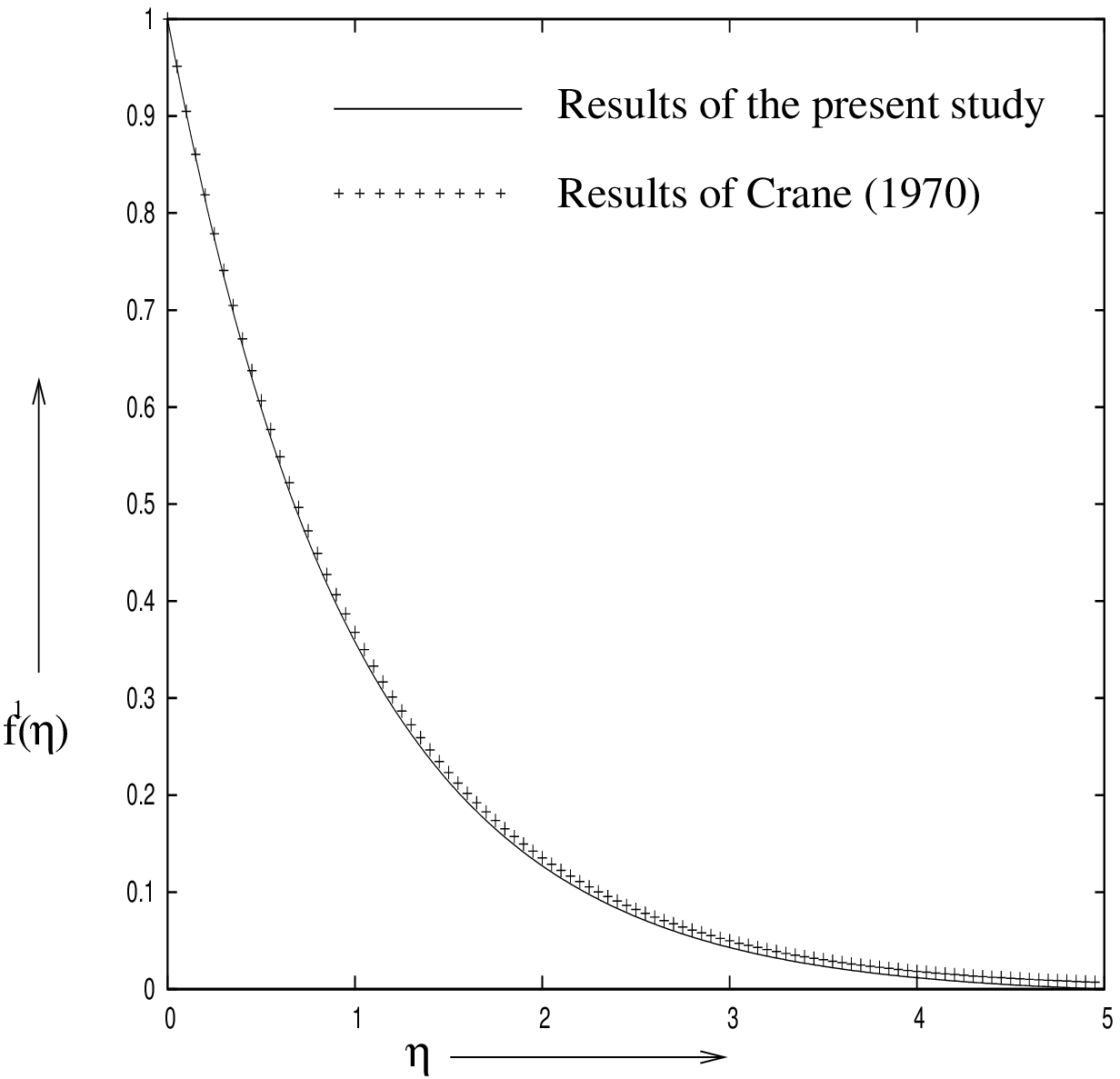}\\
Fig. 2 ~~ Comparison of axial velocity profile $f'(\eta)$ with the
results of Crane \cite{Crane} for\\
 $ M$ = $ Gr$ = $Gc$ = $Sc$ =$
Pr$ = $Nr$ =  0.0 \\
\end{center}
\end{minipage}\vspace*{.5cm}\\

\begin{minipage}{1.0\textwidth}
   \begin{center}
      \includegraphics[width=3.8in,height=2.5in ]{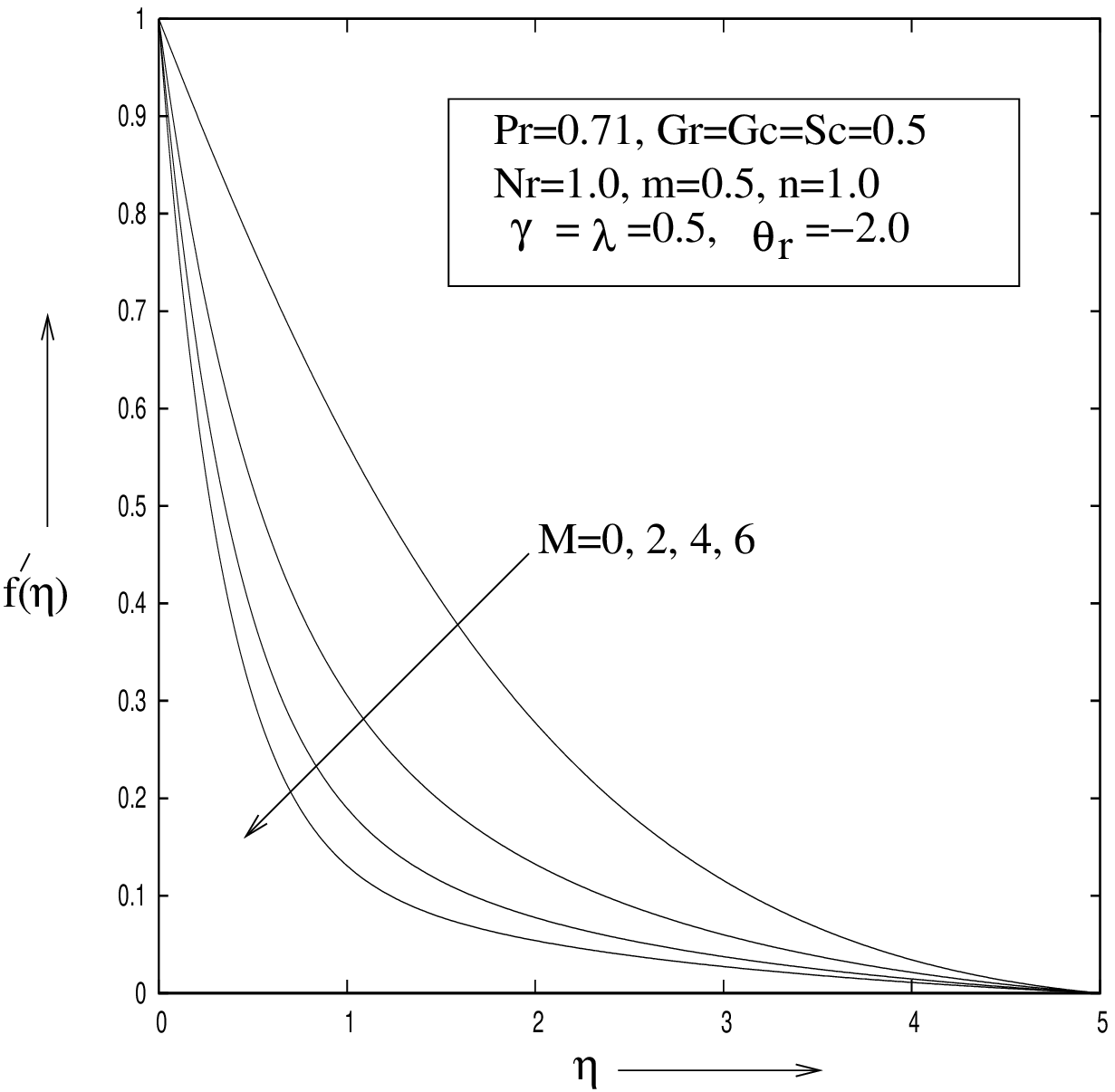}\\
Fig. 3~~ Variation of $f'(\eta)$ with $\eta$ for different values
    of~$M$\vspace*{1.5cm}  \\

  \includegraphics[width=3.8in,height=2.5in ]{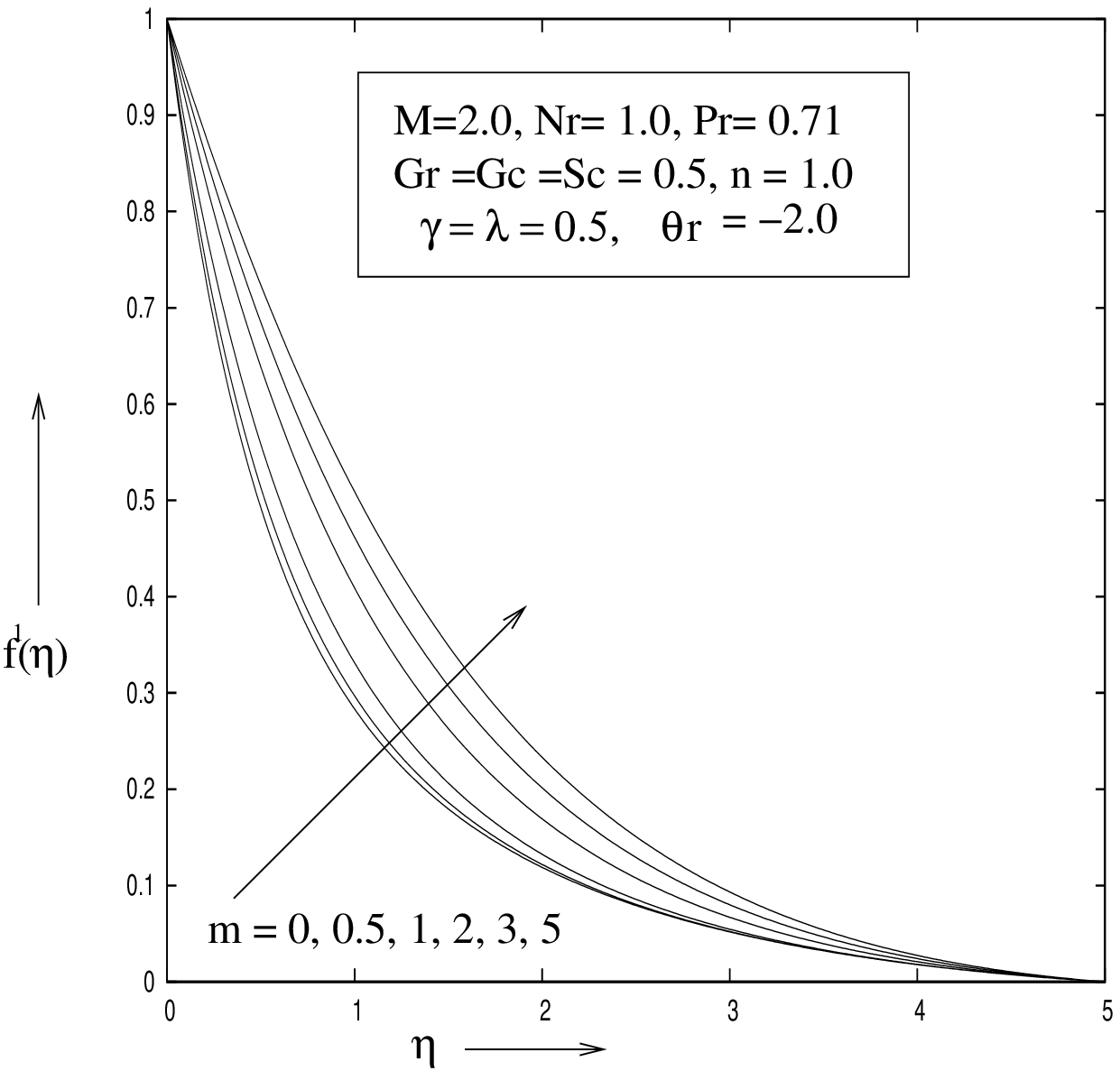}\\
Fig. 4 ~~ Variation of $f'(\eta)$ with $\eta$ for different values
    of~$ m$ \\
\end{center}
\end{minipage}\vspace*{.5cm}\\

\begin{minipage}{1.0\textwidth}
   \begin{center}
      \includegraphics[width=3.8in,height=2.5in ]{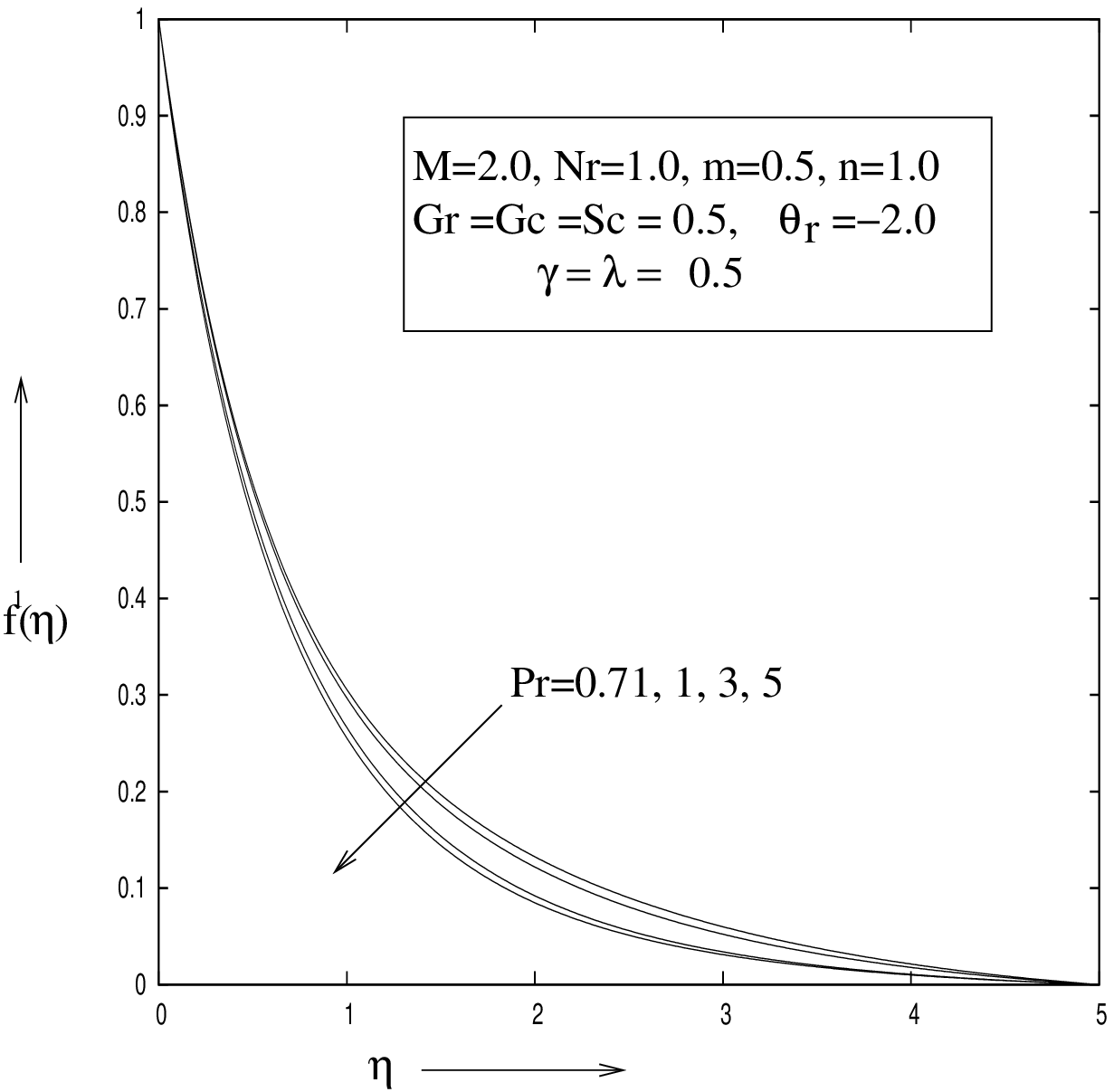}\\
Fig. 5 ~~ Variation of $f'(\eta)$ with $\eta$
 for different values of $Pr$\vspace*{1.5cm}\\

\includegraphics[width=3.8in,height=2.5in ]{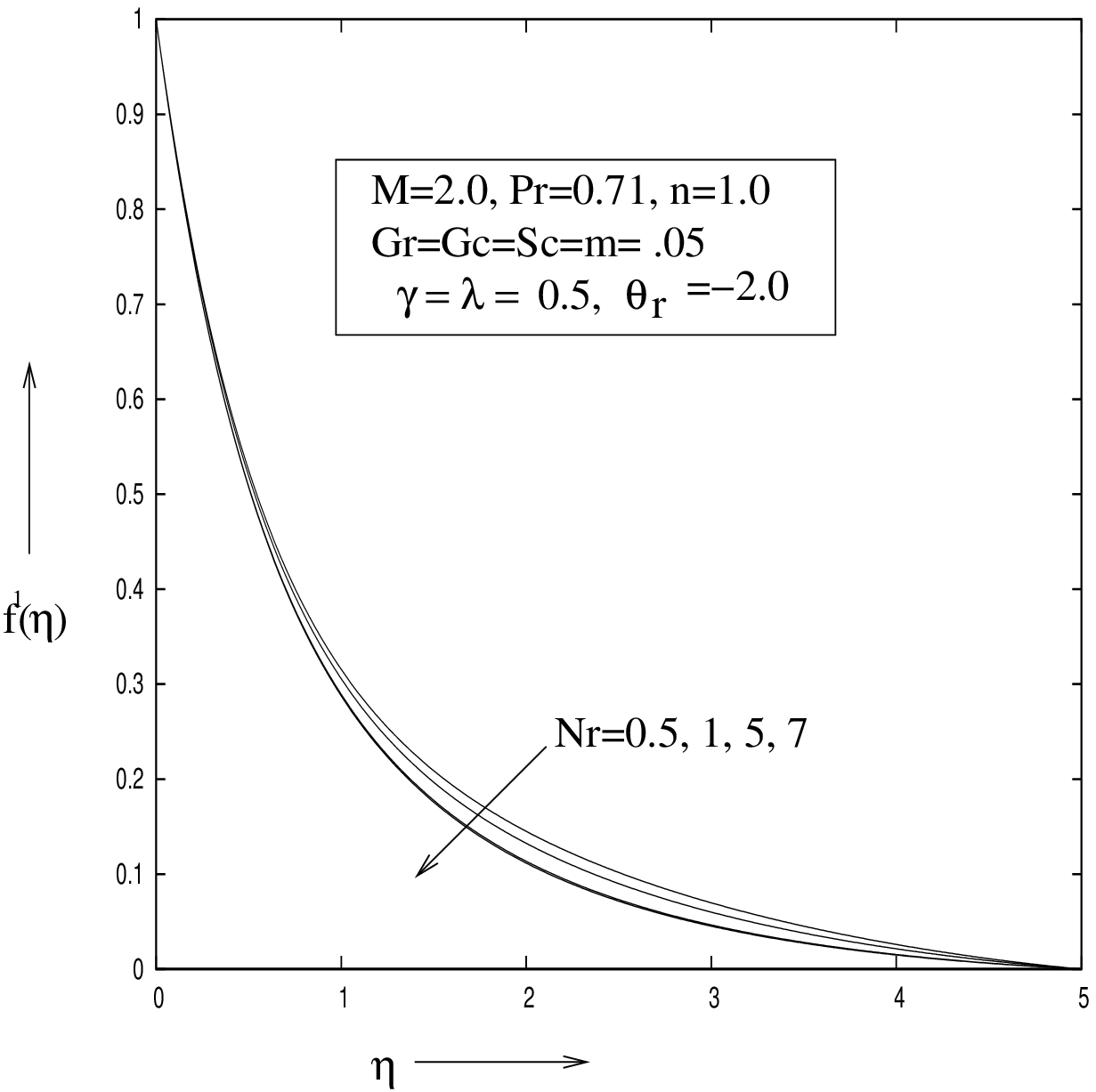}\\
Fig. 6~~  Variation of $f'(\eta)$ with $\eta$
 for different values of~$Nr$  \\

\end{center}
\end{minipage}\vspace*{.5cm}\\

\begin{minipage}{1.0\textwidth}
   \begin{center}
      \includegraphics[width=3.8in,height=2.5in ]{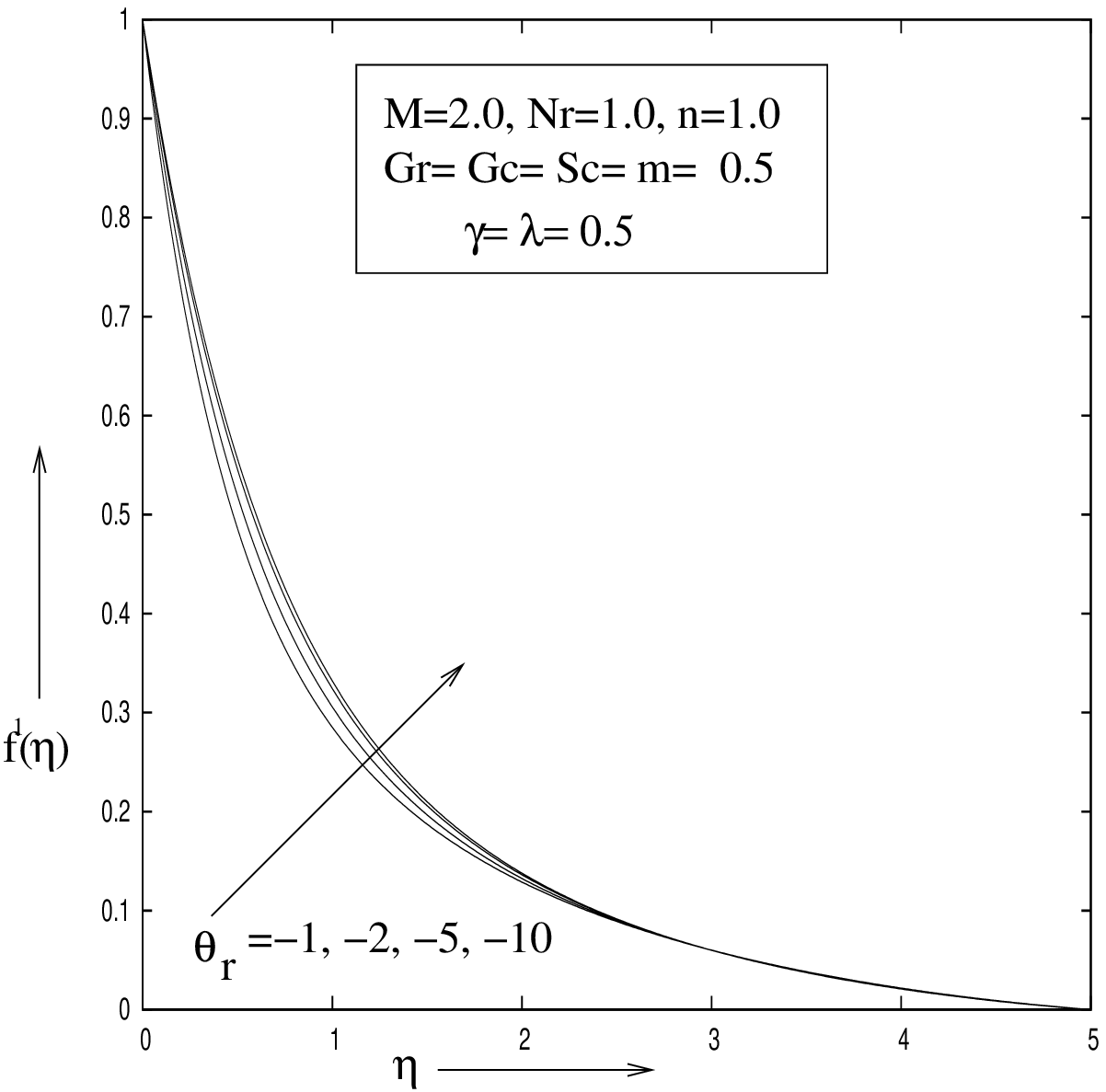}\\
Fig. 7 ~~ Variation of $f'(\eta)$ with $\eta$ for different
       values of $\theta_r$\vspace*{1.5cm} \\

\includegraphics[width=3.8in,height=2.5in ]{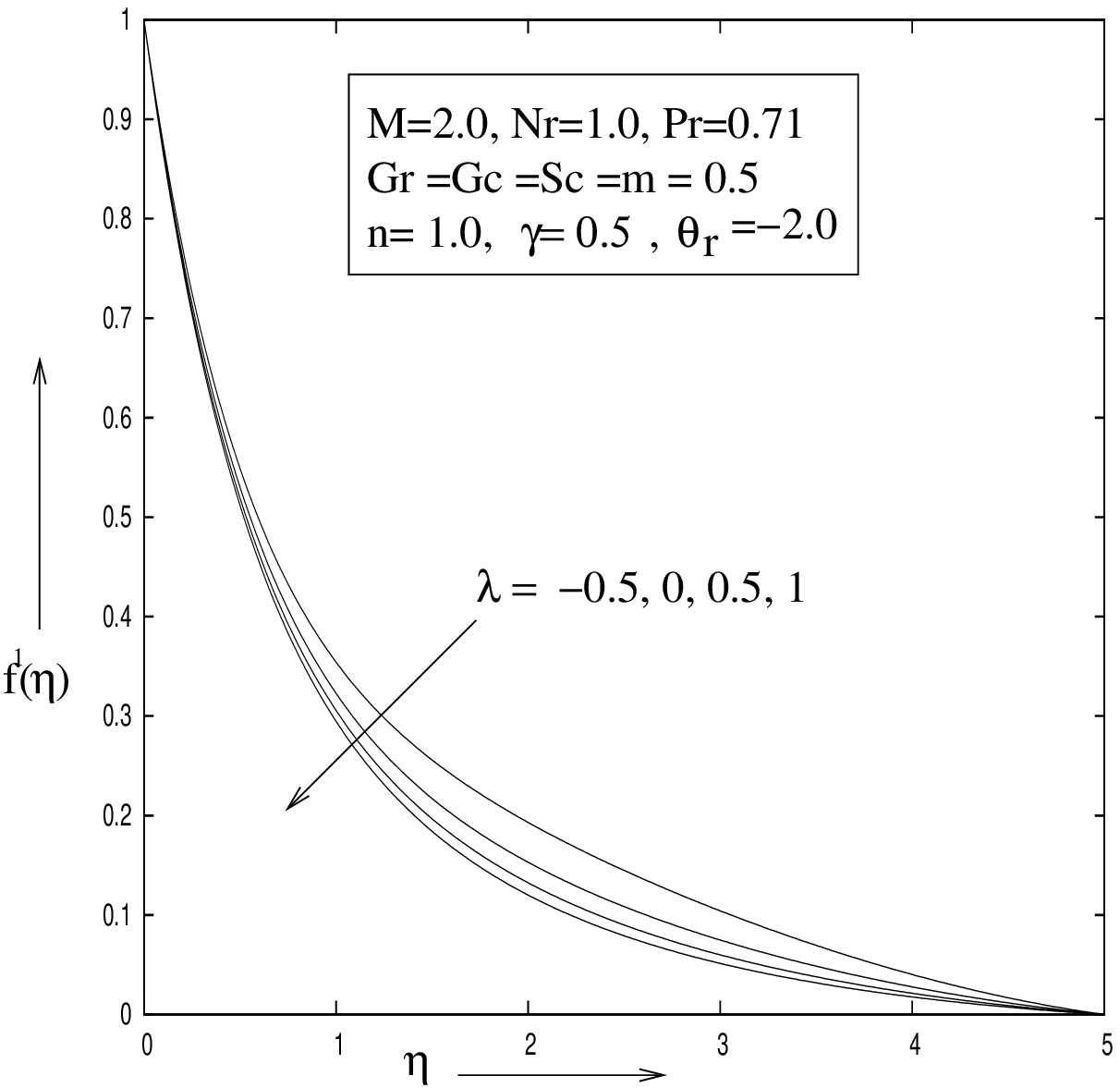}\\
Fig. 8 ~~ Variation of $f'(\eta)$ with $\eta$ for different
      values of $\lambda$ \\
    \end{center}
    \end{minipage}\vspace*{.5cm}\\

\begin{minipage}{1.0\textwidth}
   \begin{center}
      \includegraphics[width=3.8in,height=2.5in ]{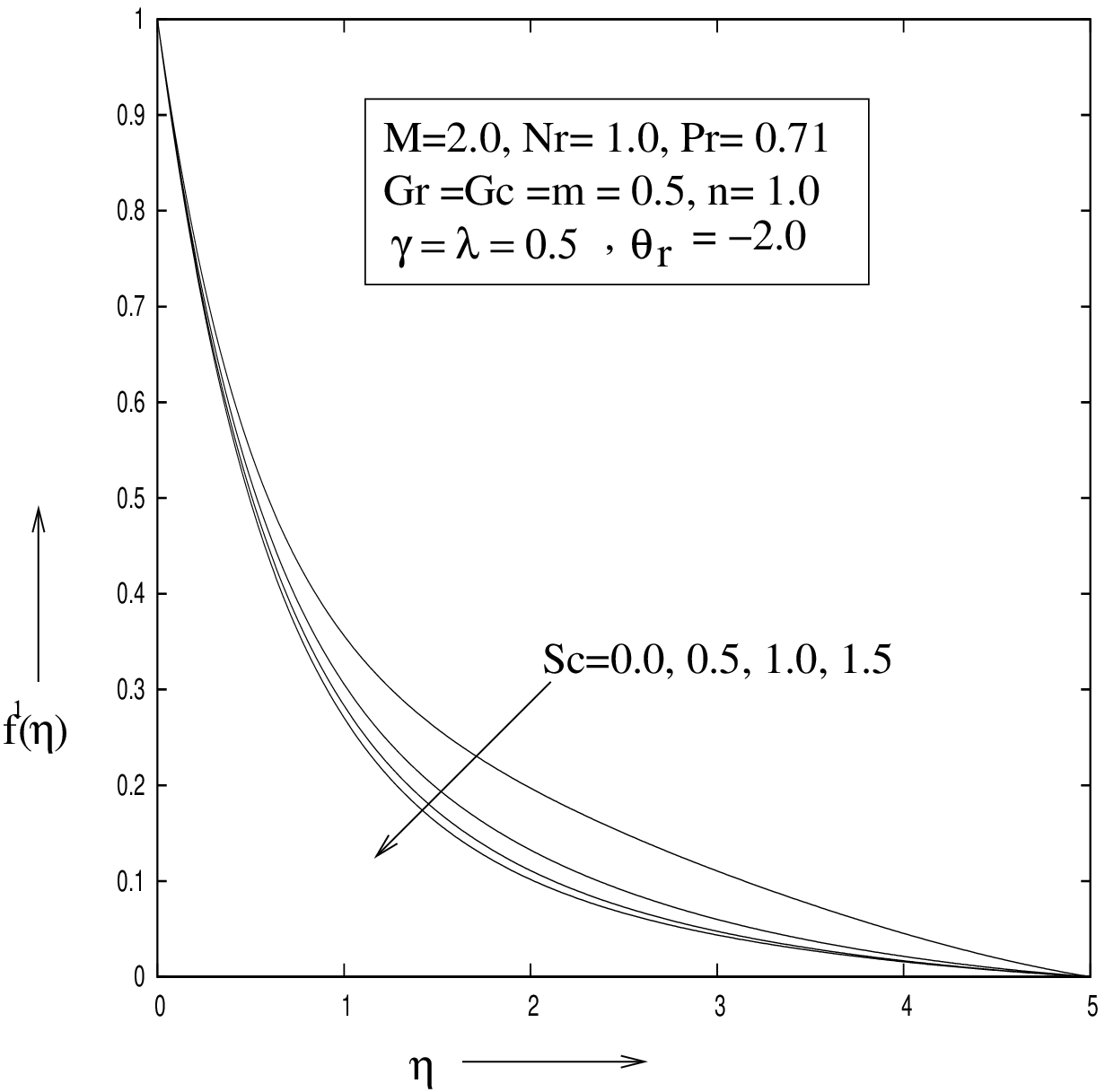}\\
Fig. 9 ~~  Variation of $f'(\eta)$ with $\eta$ for different
       values of $Sc$\vspace*{1.5cm} \\

      \includegraphics[width=3.8in,height=2.5in ]{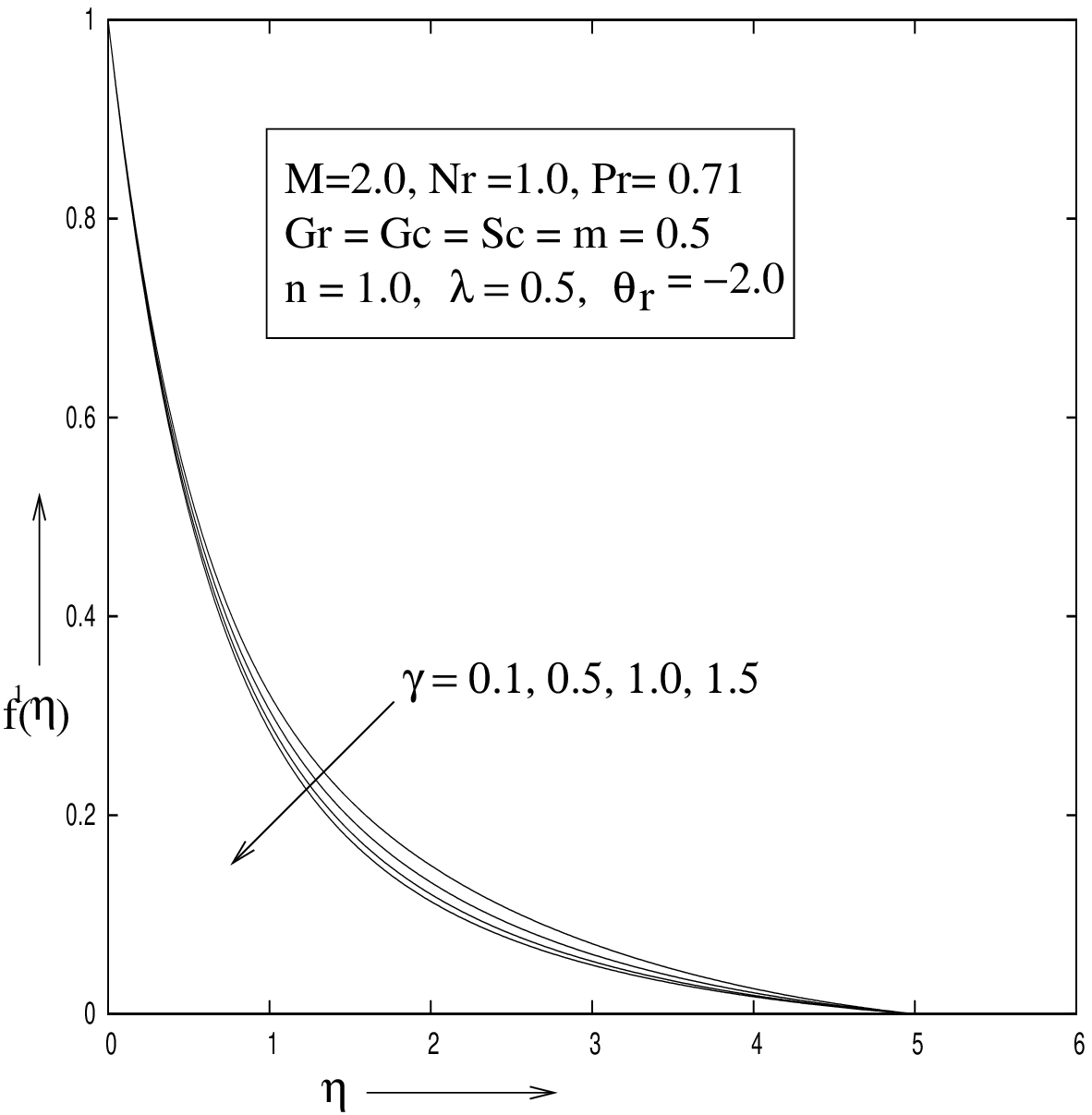}\\
Fig. 10~~  Variation of $f'(\eta)$ with $\eta$ for different
       values of $\gamma$ \\
\end{center}
\end{minipage}\vspace*{.5cm}\\

\begin{minipage}{1.0\textwidth}
   \begin{center}
      \includegraphics[width=3.8in,height=2.5in ]{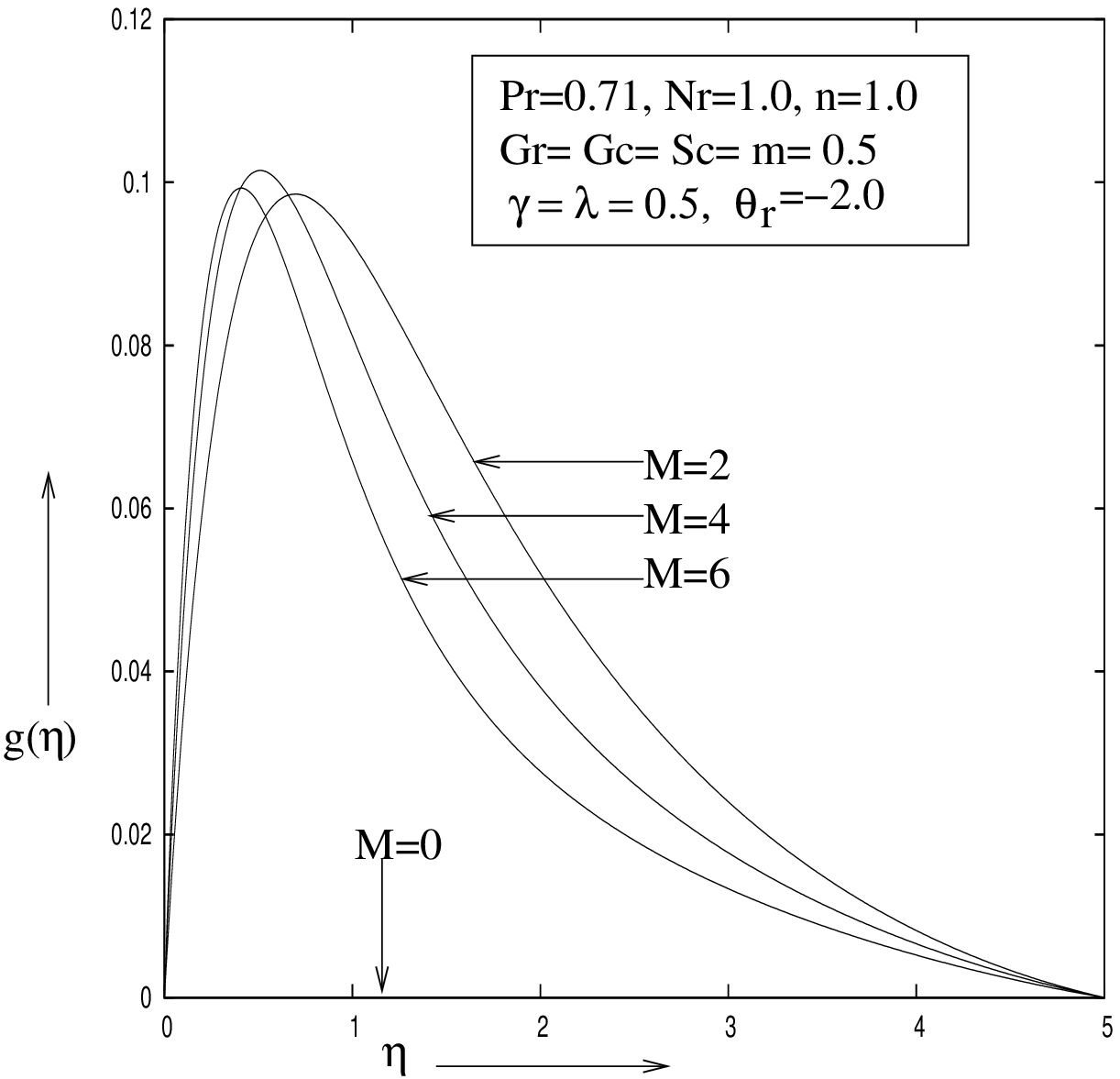}\\
Fig. 11 ~~ Variation of $g(\eta)$ with $\eta$ for different values
    of~$ M$\vspace*{1.5cm} \\

\includegraphics[width=3.8in,height=2.5in ]{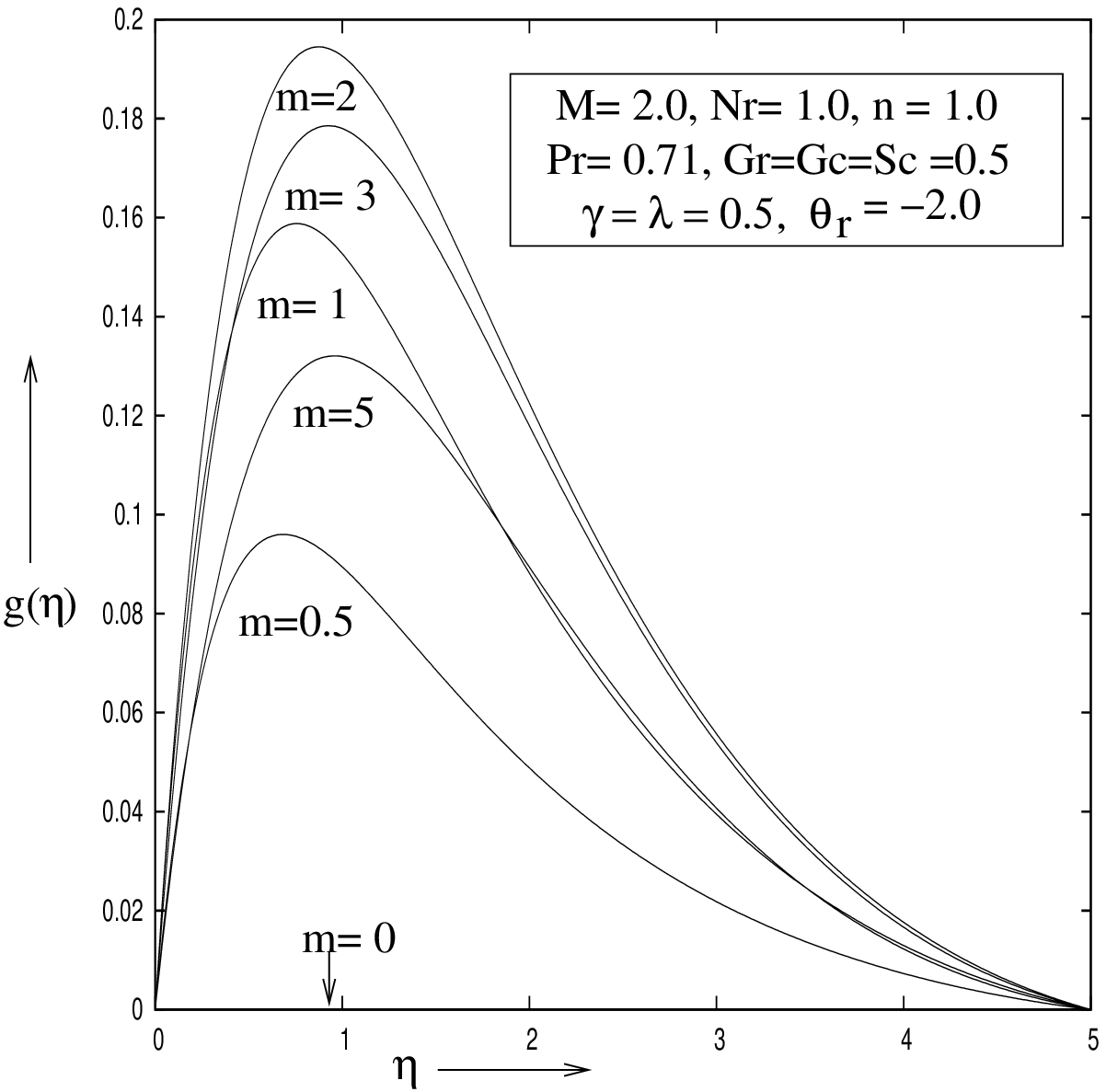}\\
Fig. 12~~Variation of $g(\eta)$ for different values of the Hall
parameter $m$.
  \end{center}
\end{minipage}\vspace*{.5cm}\\

   \begin{minipage}{1.0\textwidth}
   \begin{center}
      \includegraphics[width=3.8in,height=2.5in ]{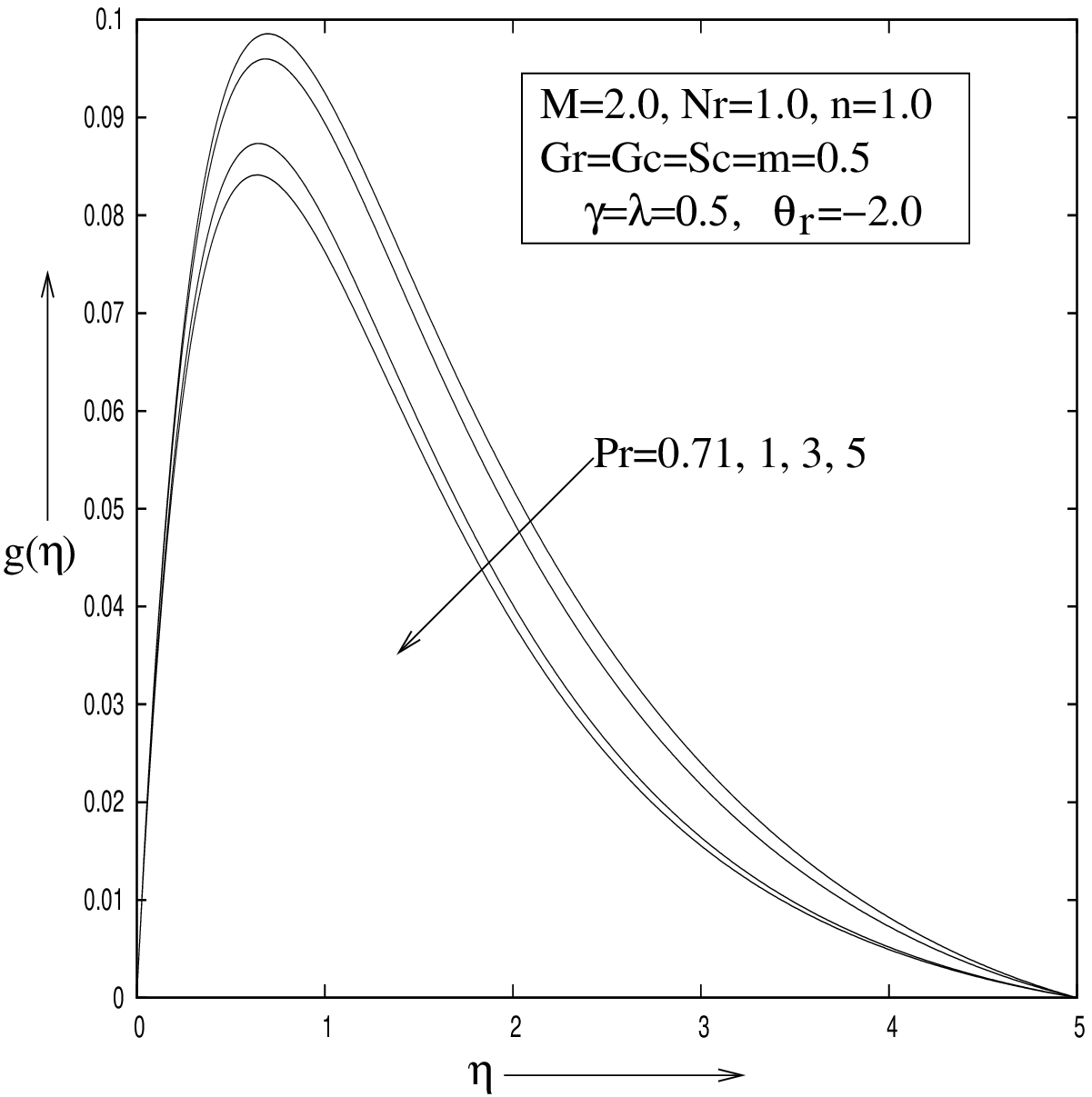}\\
Fig. 13~~Variation of $g(\eta)$ with $\eta$ for different
        values of $ Pr $\vspace*{1.5cm} \\

\includegraphics[width=3.8in,height=2.5in ]{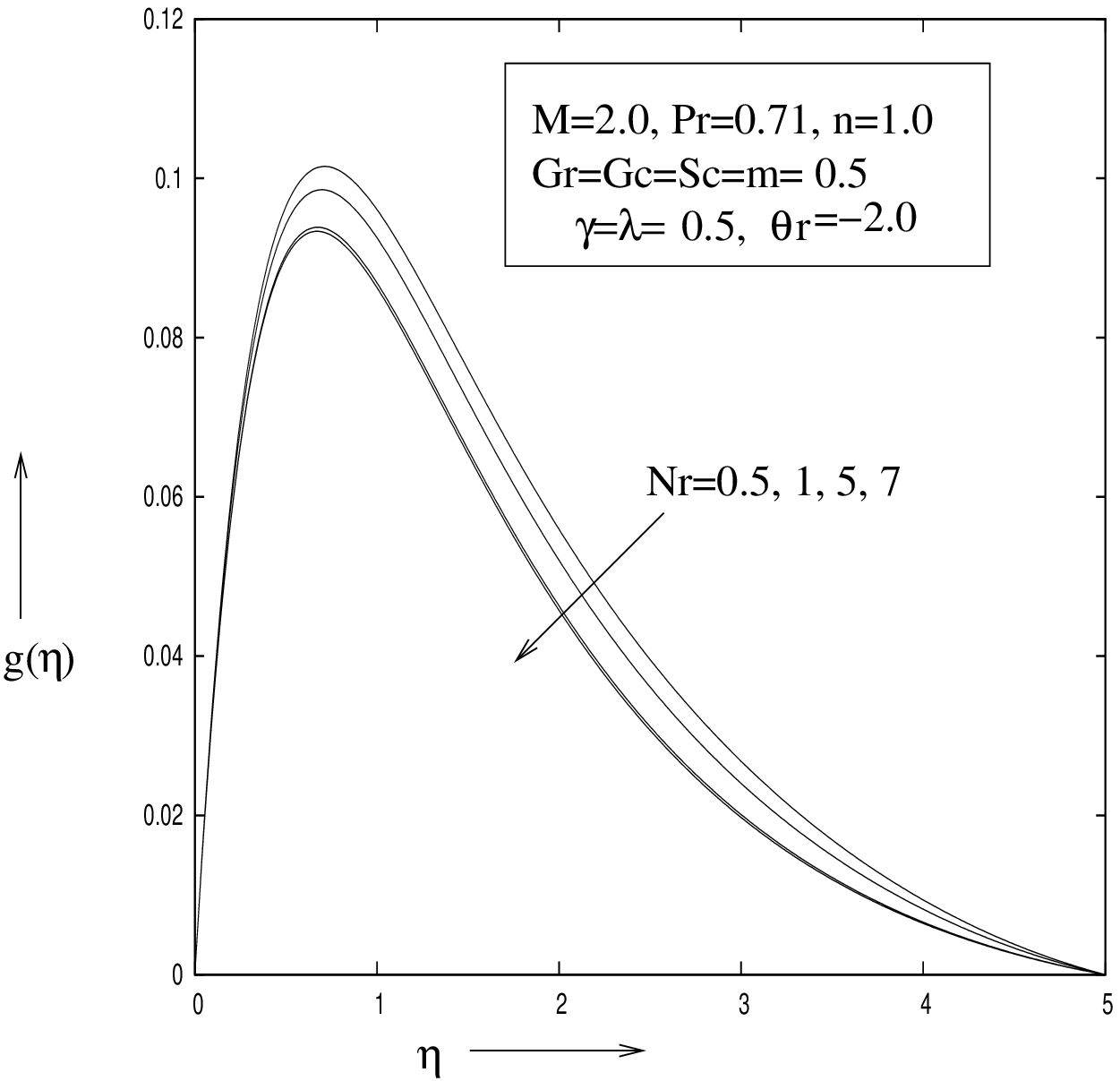}\\
Fig. 14~~Variation of $g(\eta)$ with $\eta$ for different
       values of~$Nr$  \\
        \end{center}
      \end{minipage}\vspace*{.5cm}\\

\begin{minipage}{1.0\textwidth}
   \begin{center}
      \includegraphics[width=3.8in,height=2.5in ]{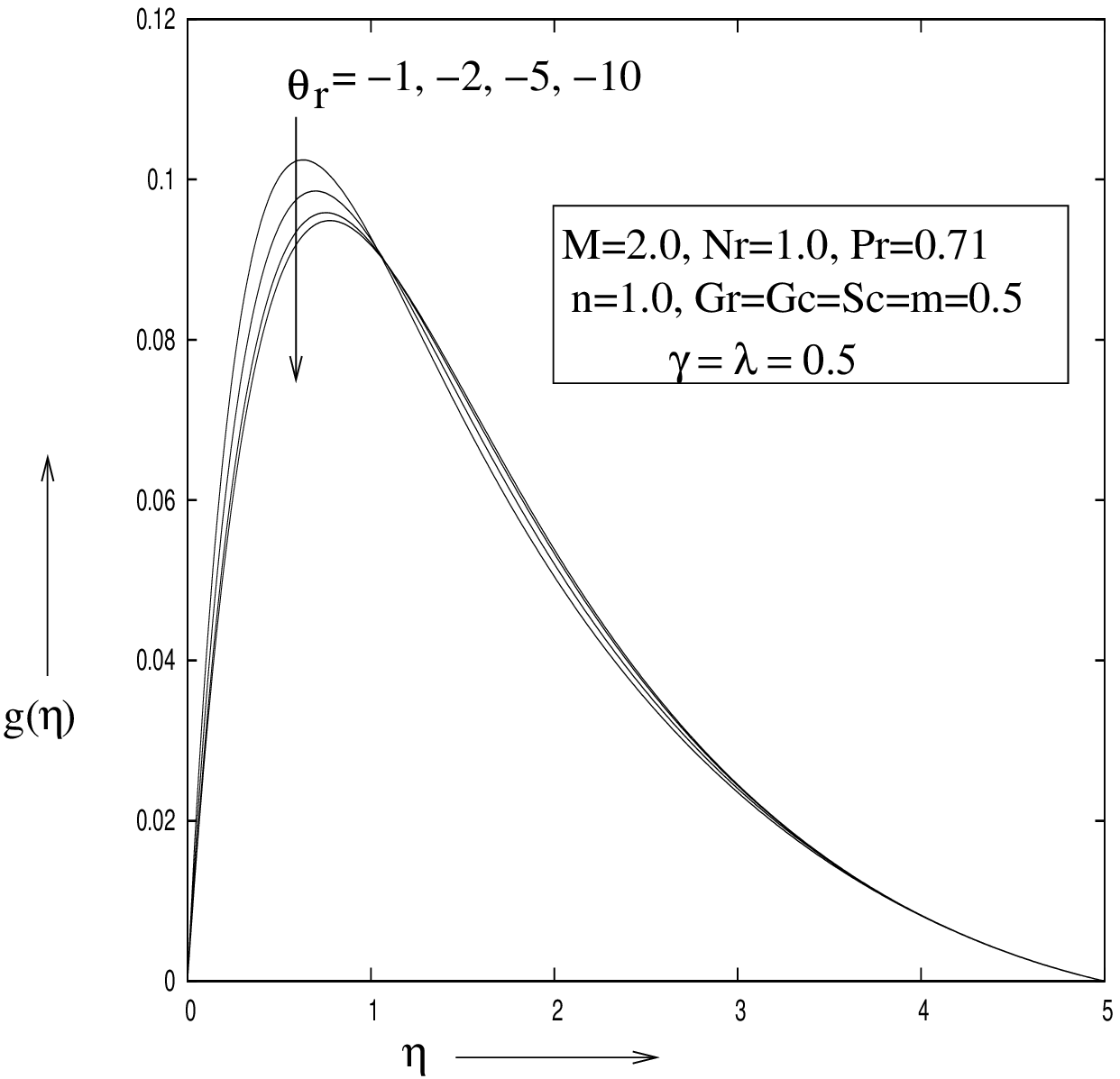}\\
Fig. 15~~  Variation of $g(\eta)$ with $\eta$
 for different values of~ $\theta_r$\vspace*{1.5cm}  \\

      \includegraphics[width=3.8in,height=2.5in ]{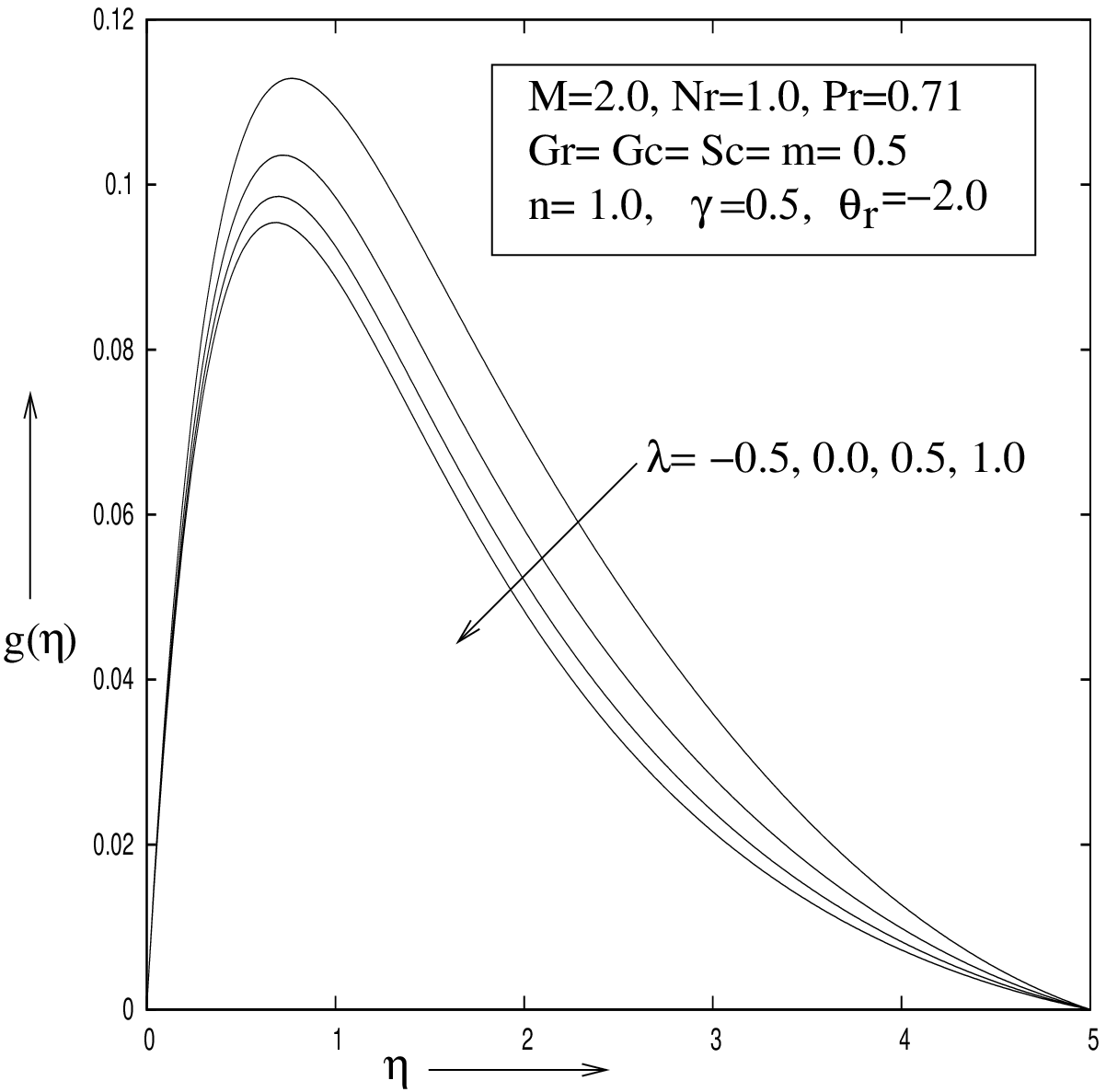}\\
Fig. 16~~ Variation of $g(\eta)$ with $\eta$ for different
       values of~$\lambda$  \\

  \end{center}
      \end{minipage}\vspace*{.5cm}\\

\begin{minipage}{1.0\textwidth}
   \begin{center}
\includegraphics[width=3.8in,height=2.5in ]{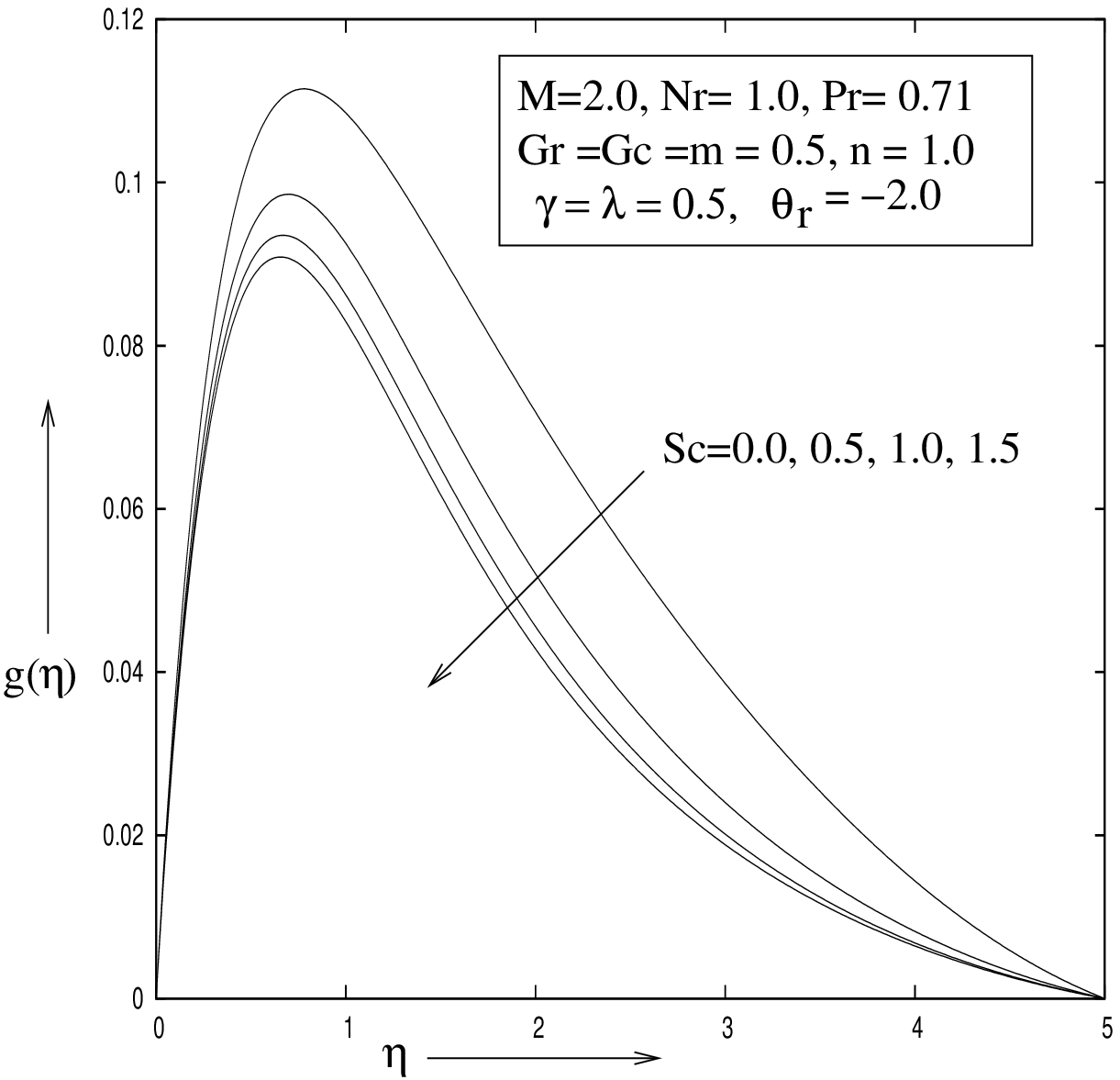}\\
Fig. 17~~Variation of $g(\eta)$ with $\eta$ for different
       values of~$Sc$\vspace*{1.5cm}  \\

\includegraphics[width=3.8in,height=2.5in ]{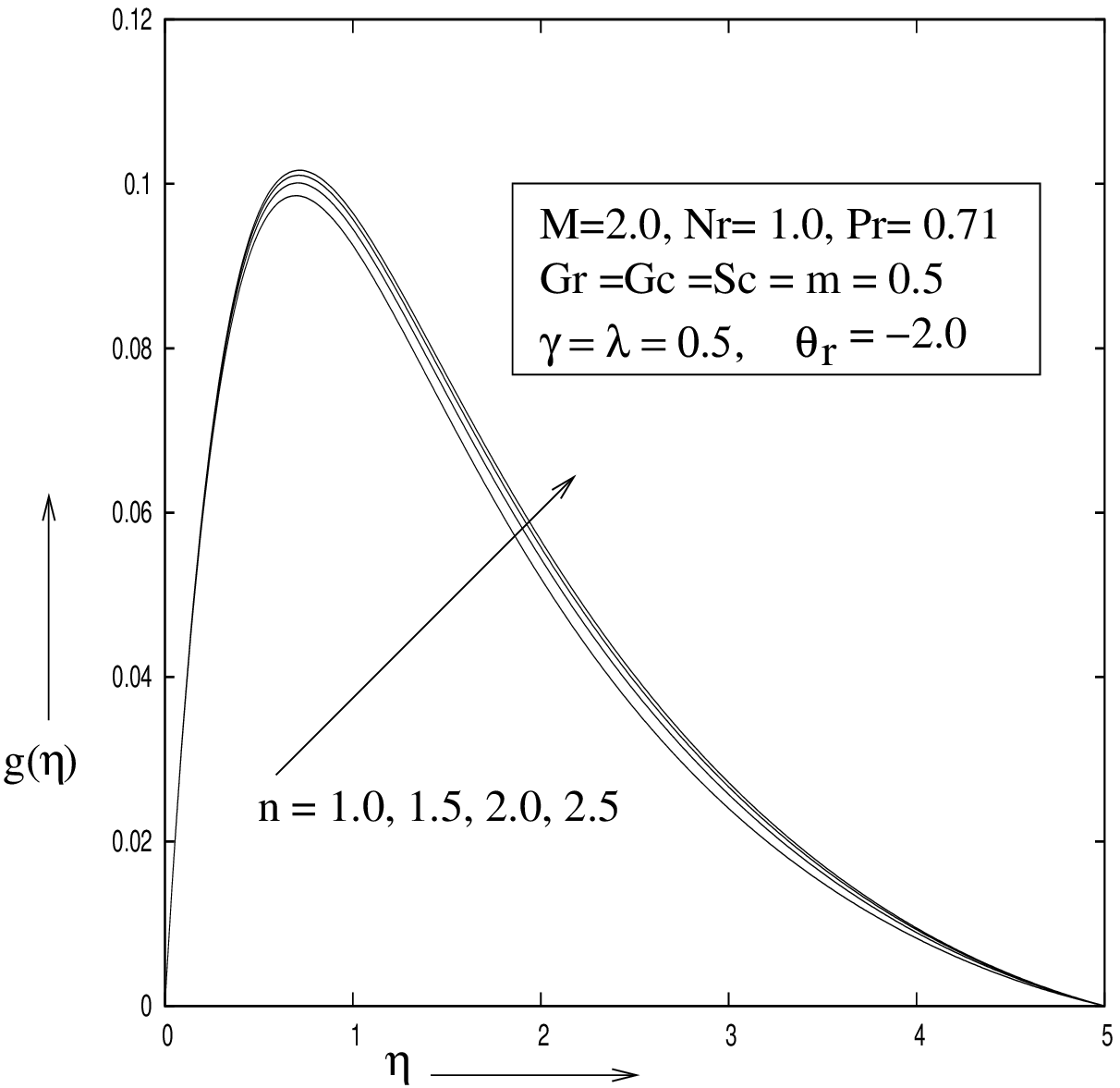}\\
Fig. 18~~Variation of $g(\eta)$ with $\eta$ for different
       values of~$n$  \\

\end{center}
\end{minipage}\vspace*{.5cm}\\

\begin{minipage}{1.0\textwidth}
   \begin{center}
\includegraphics[width=3.8in,height=2.5in ]{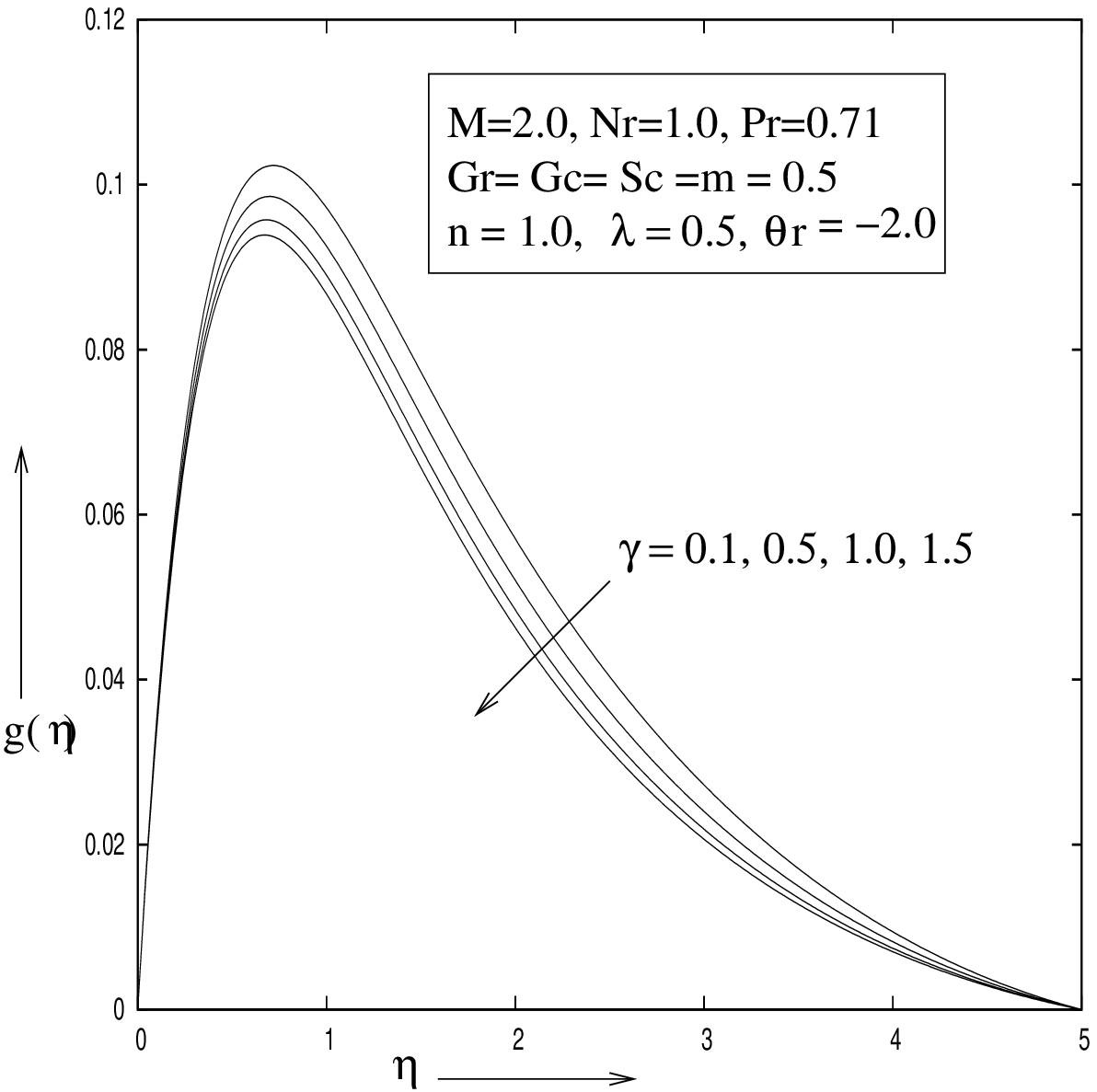}\\
Fig. 19 ~~ Variation of $g(\eta)$ with $\eta$ for different
       values of~$\gamma$\vspace*{1.5cm}  \\

\includegraphics[width=3.8in,height=2.5in ]{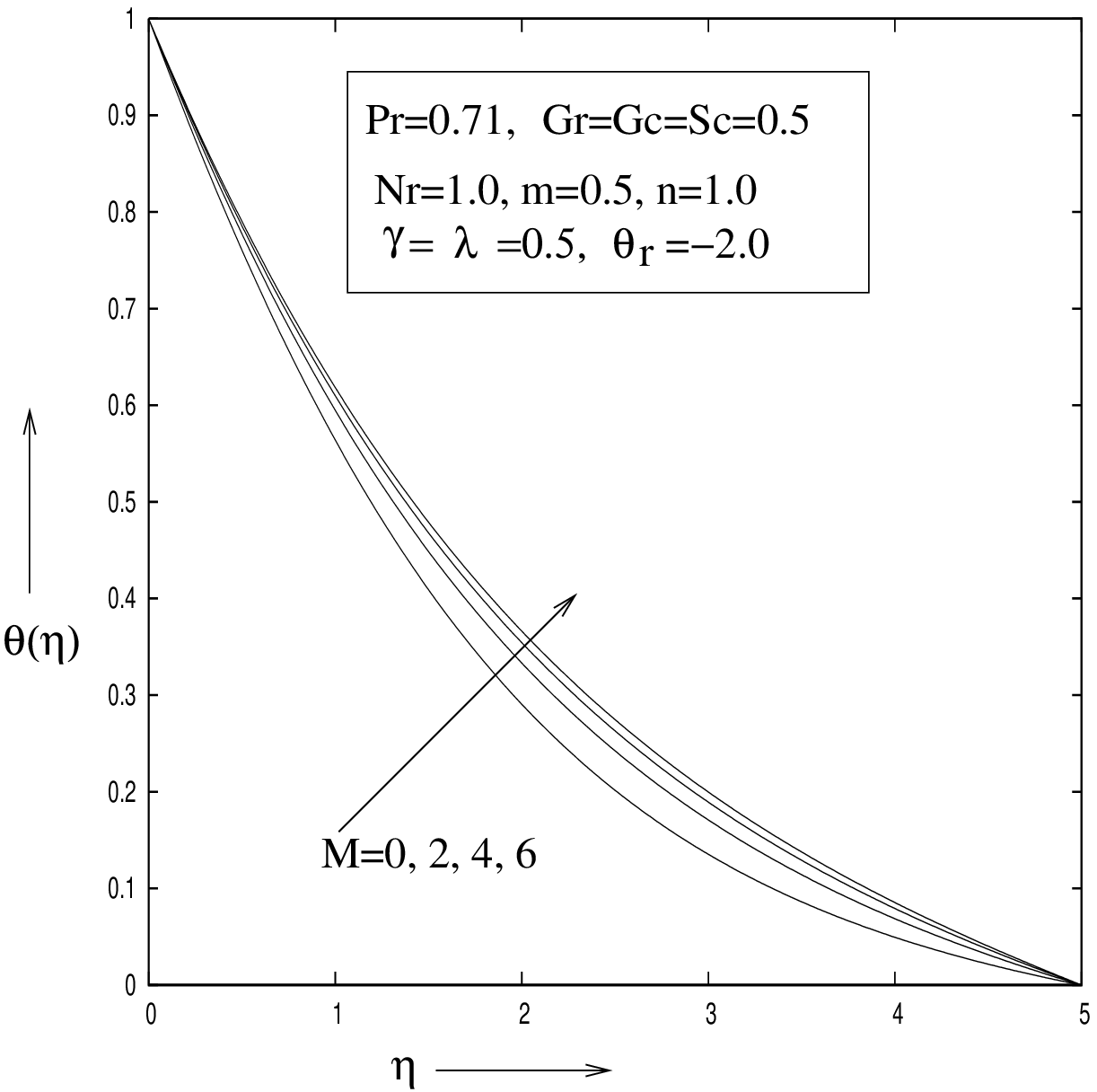}\\
Fig. 20~~  Distribution of dimensionless temperature
$\theta(\eta)$ with $\eta$ for different
       values of~$ M$  \\
\end{center}
\end{minipage}\vspace*{.5cm}\\

\begin{minipage}{1.0\textwidth}
   \begin{center}
   \includegraphics[width=3.8in,height=2.5in ]{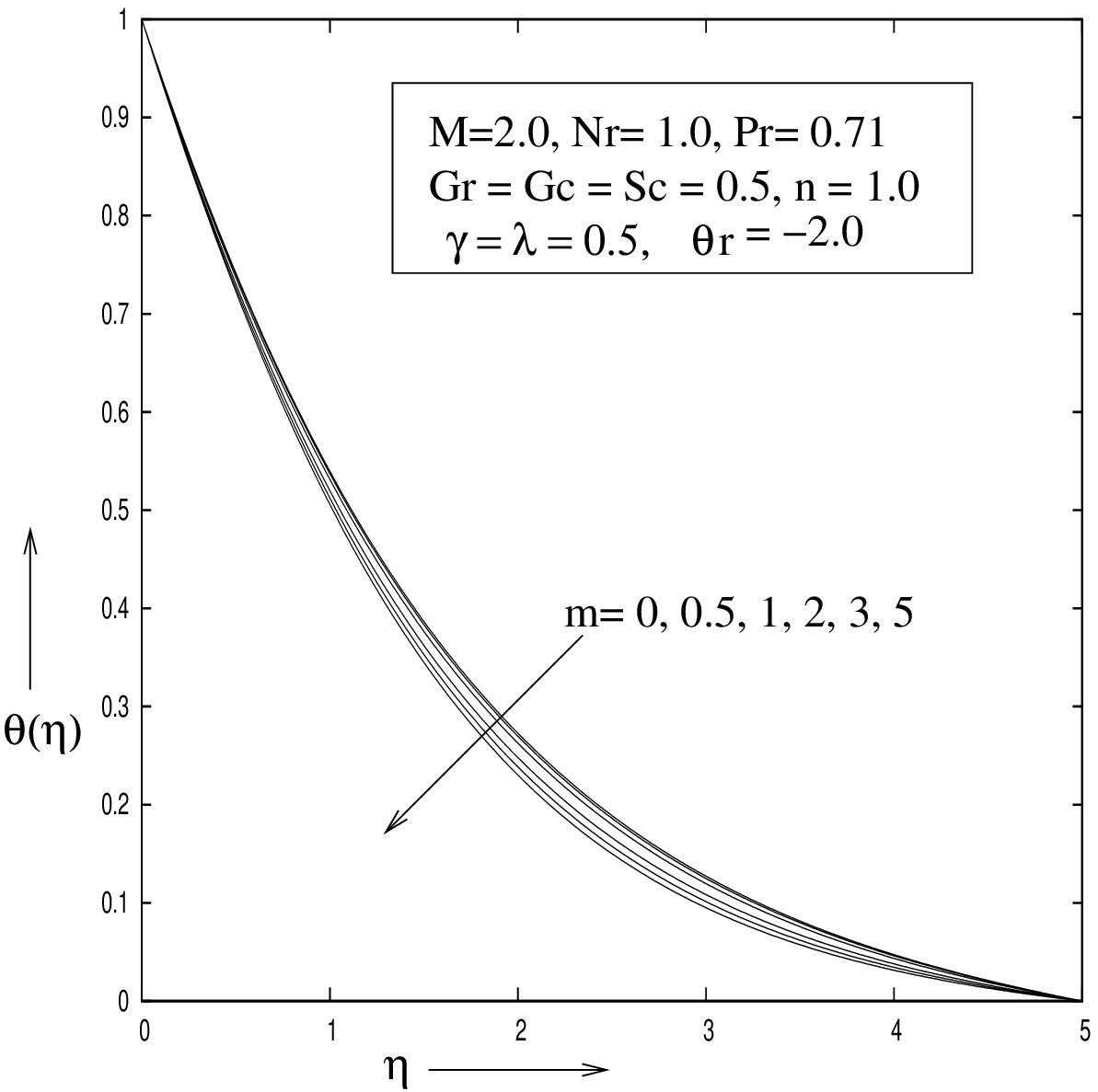}\\
 Fig. 21~~  Distribution of dimensionless temperature
$\theta(\eta)$ with $\eta$ for different
      values of $m$\vspace*{1.5cm} \\

\includegraphics[width=3.8in,height=2.5in ]{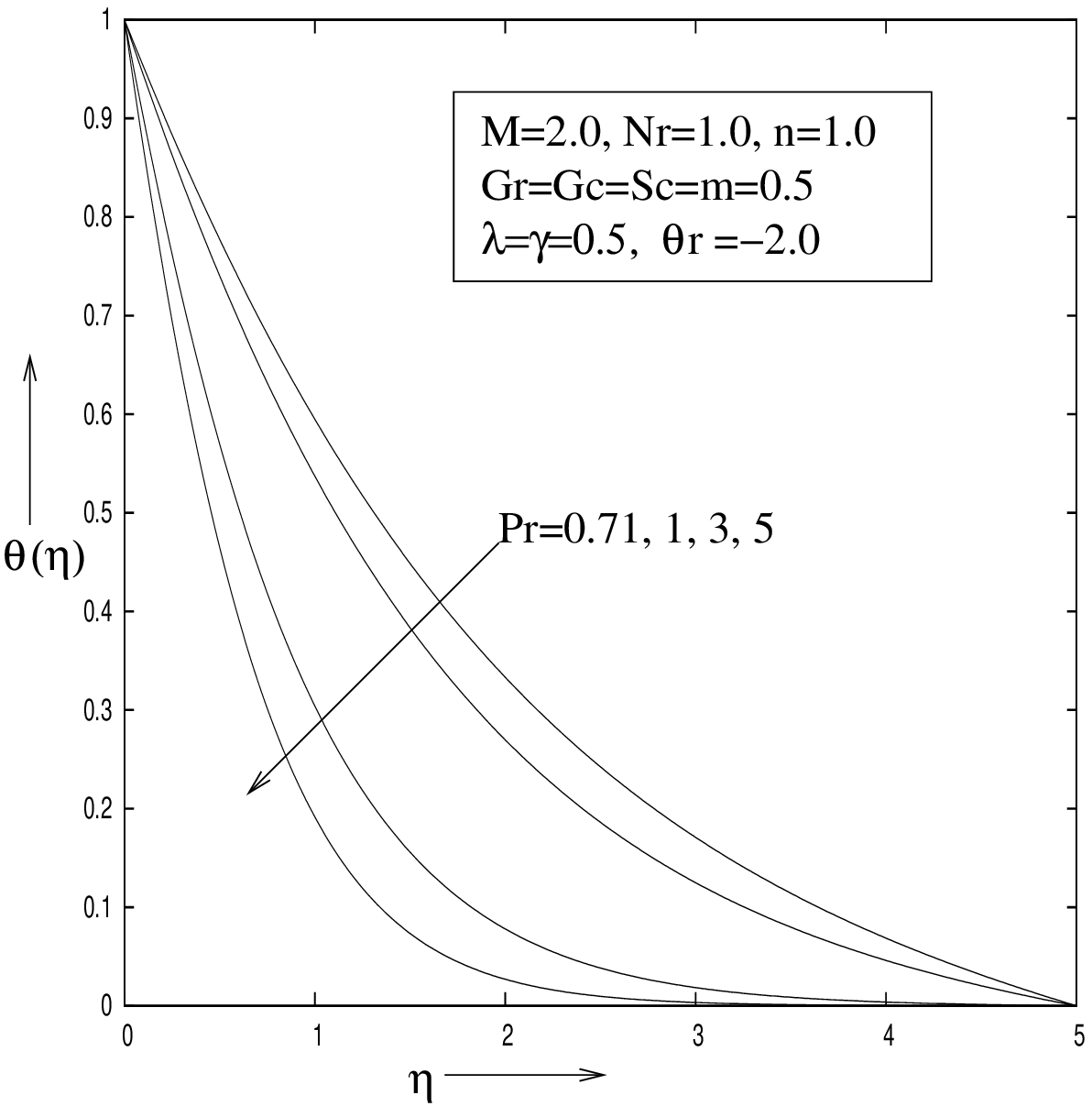}\\
Fig. 22~~Distribution of dimensionless temperature $\theta(\eta)$
with $\eta$ for different
        values of $Pr$ \\
\end{center}
\end{minipage}\vspace*{.5cm}\\

\begin{minipage}{1.0\textwidth}
   \begin{center}
\includegraphics[width=3.8in,height=2.5in ]{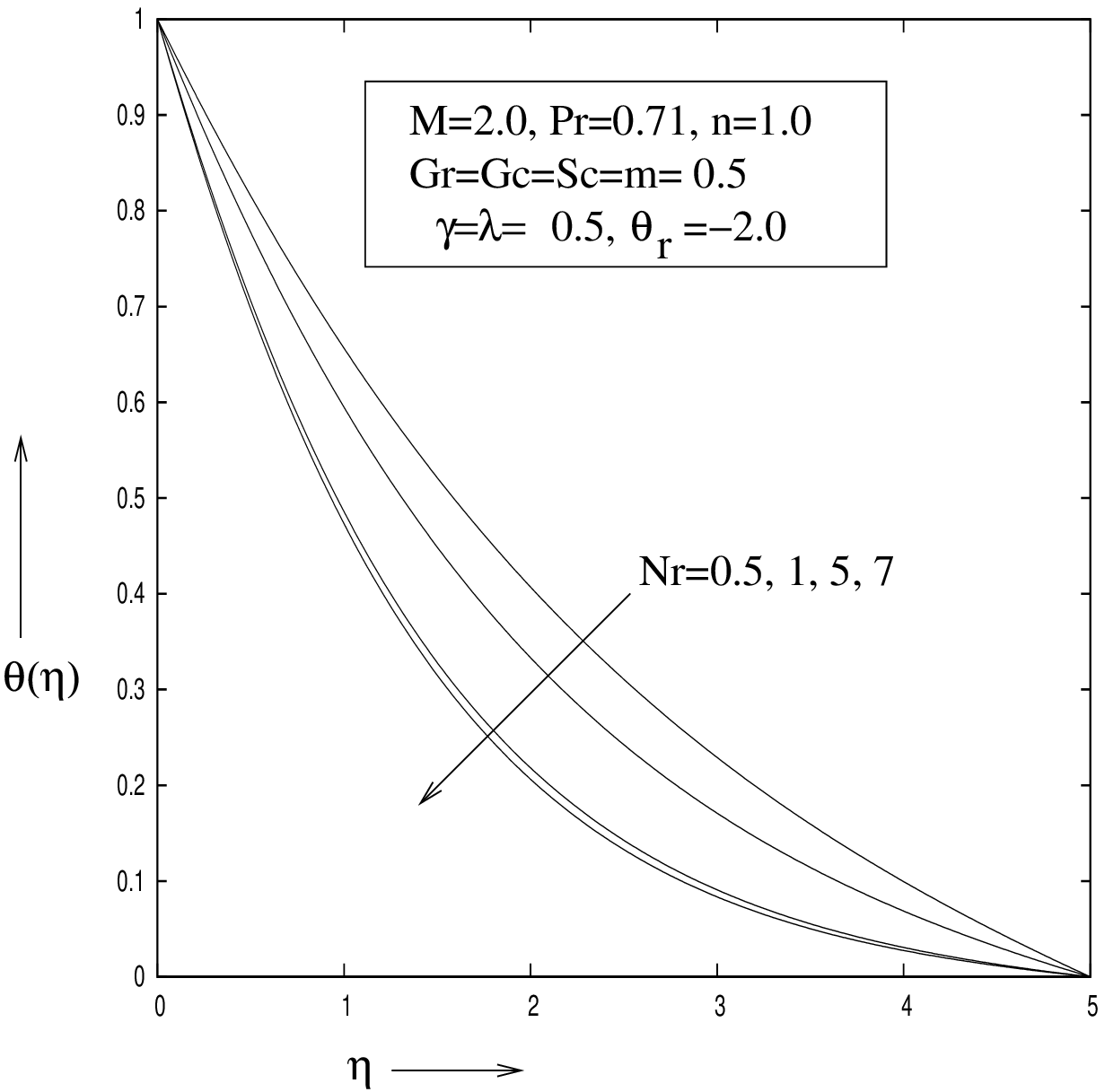}\\
Fig. 23~~ Distribution of dimensionless temperature $\theta(\eta)$
with $\eta$ for different
         values of~$Nr$\vspace*{1.5cm} \\

\includegraphics[width=3.8in,height=2.5in ]{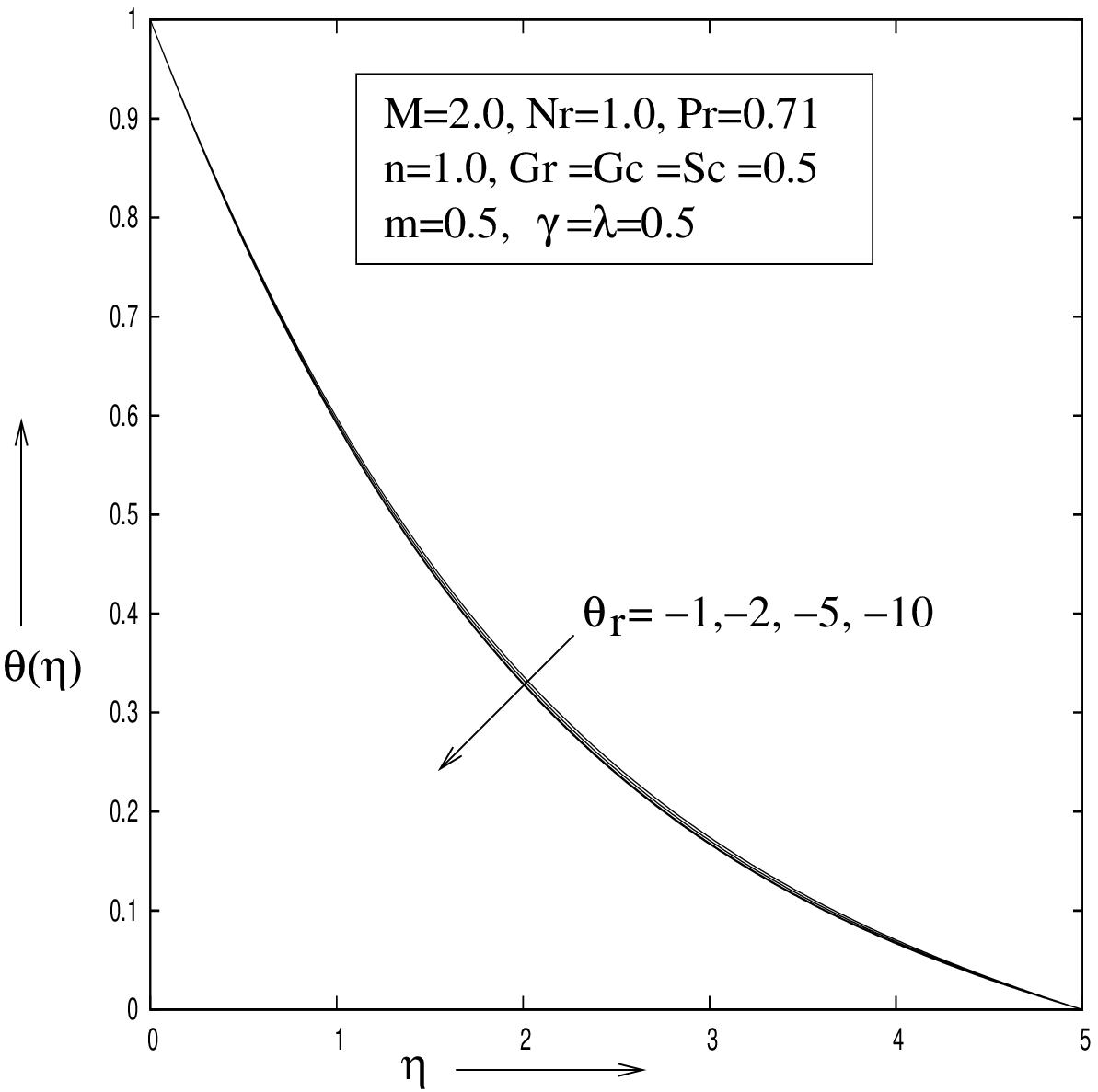}\\
Fig. 24~~ Distribution of dimensionless temperature $\theta(\eta)$
with $\eta$ for different
       values of~$\theta_r$  \\
\end{center}
\end{minipage}\vspace*{.5cm}\\

\begin{minipage}{1.0\textwidth}
   \begin{center}
\includegraphics[width=3.8in,height=2.5in ]{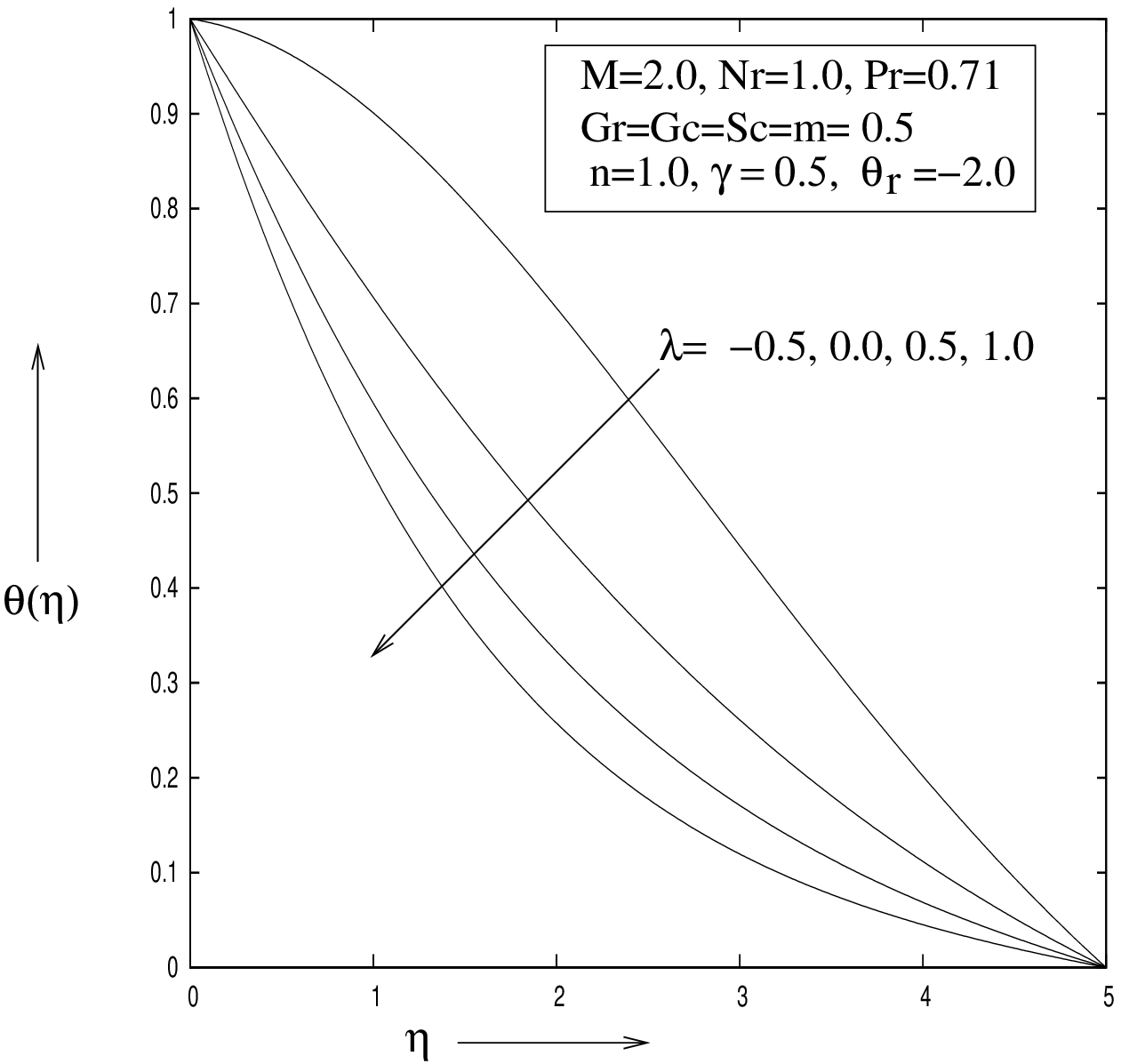}\\
Fig. 25 ~~Distribution of dimensionless temperature $\theta(\eta)$
with $\eta$ for different
       values of~$ \lambda$\vspace*{1.5cm}  \\

\includegraphics[width=3.8in,height=2.5in ]{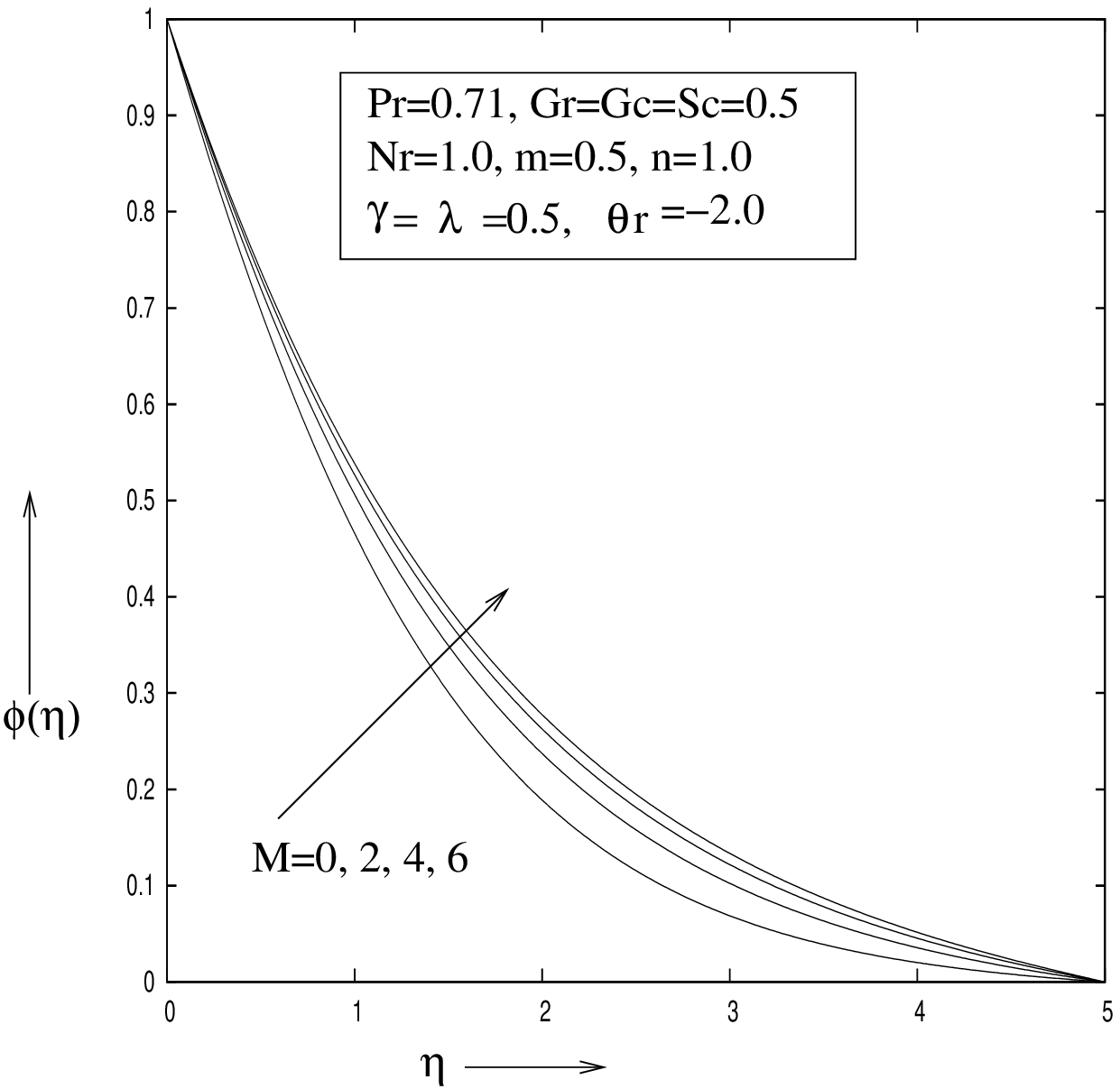}\\
Fig. 26~~  Influence of concentration species $\phi(\eta)$ with
$\eta$ for different
       values of~$M$  \\
\end{center}
\end{minipage}\vspace*{.5cm}\\

\begin{minipage}{1.0\textwidth}
   \begin{center}
   \includegraphics[width=3.8in,height=2.5in ]{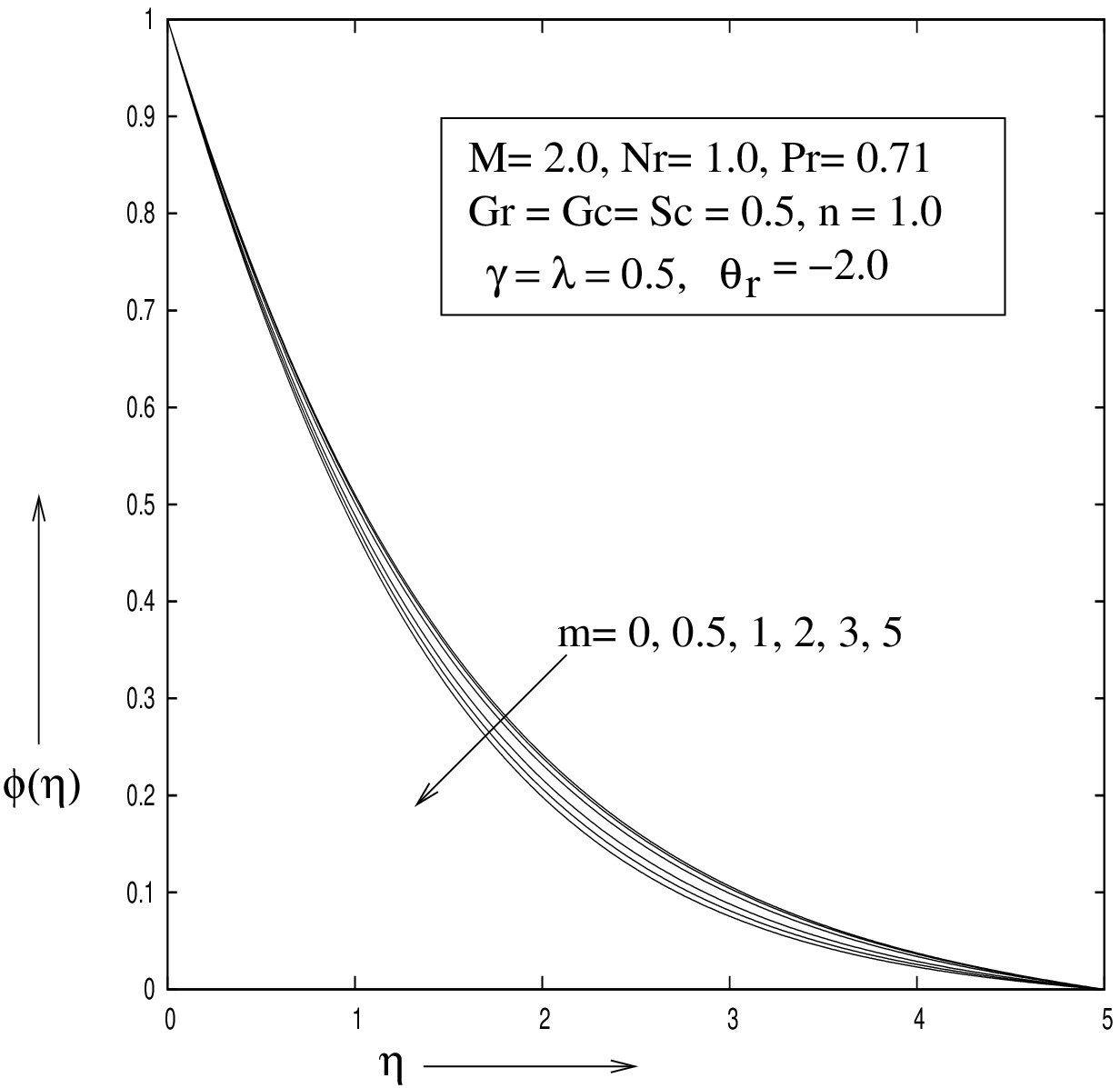}\\
 Fig. 27~~Influence of concentration species $\phi(\eta)$ with
$\eta$ for different values of~$m $\vspace*{1.5cm} \\

\includegraphics[width=3.8in,height=2.5in ]{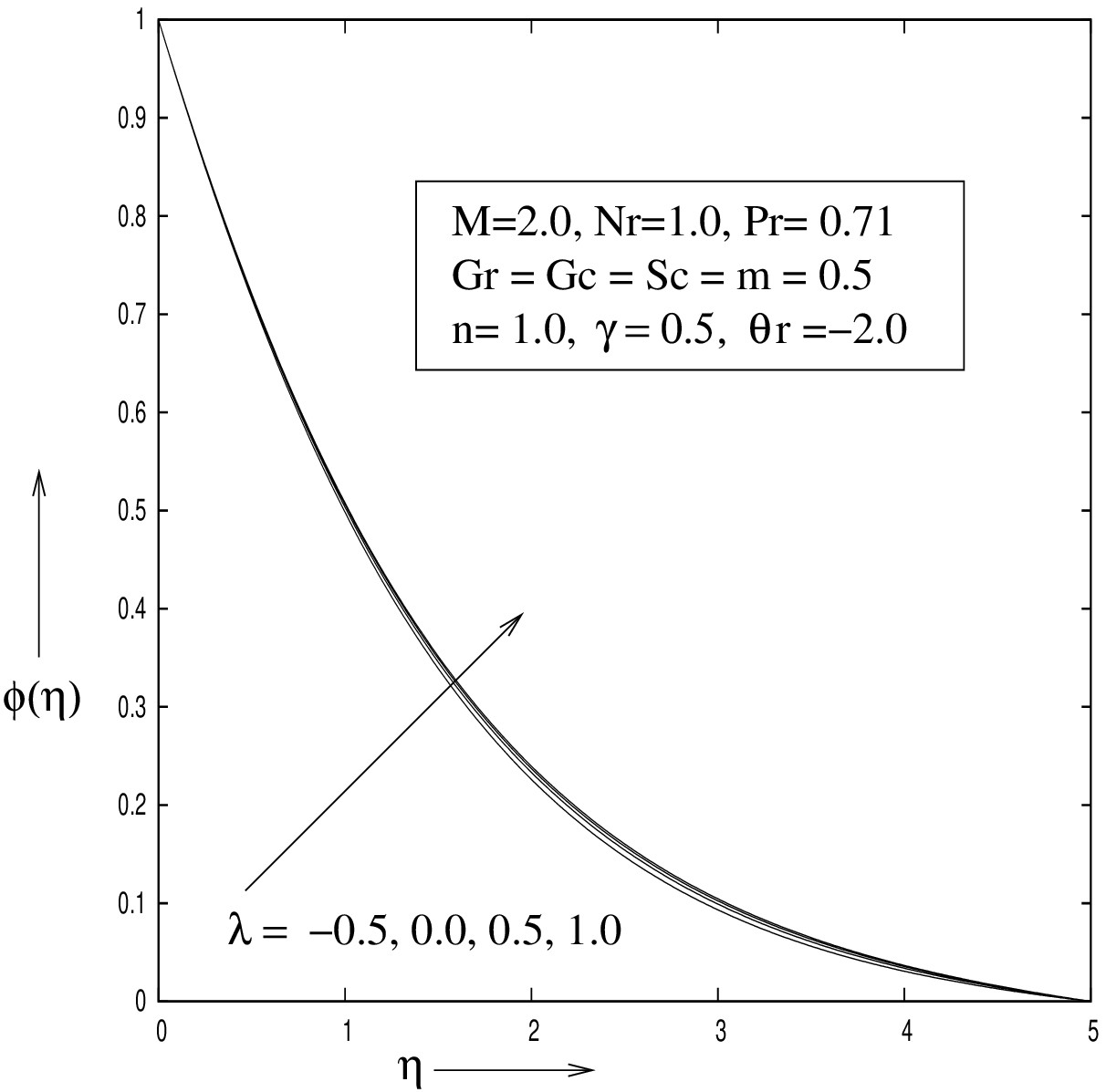}\\
Fig. 28~~ Influence of concentration species $\phi(\eta)$ with
$\eta$ for different
       values of $\lambda$ \\
\end{center}
\end{minipage}\vspace*{.5cm}\\

\begin{minipage}{1.0\textwidth}
   \begin{center}
\includegraphics[width=3.8in,height=2.5in ]{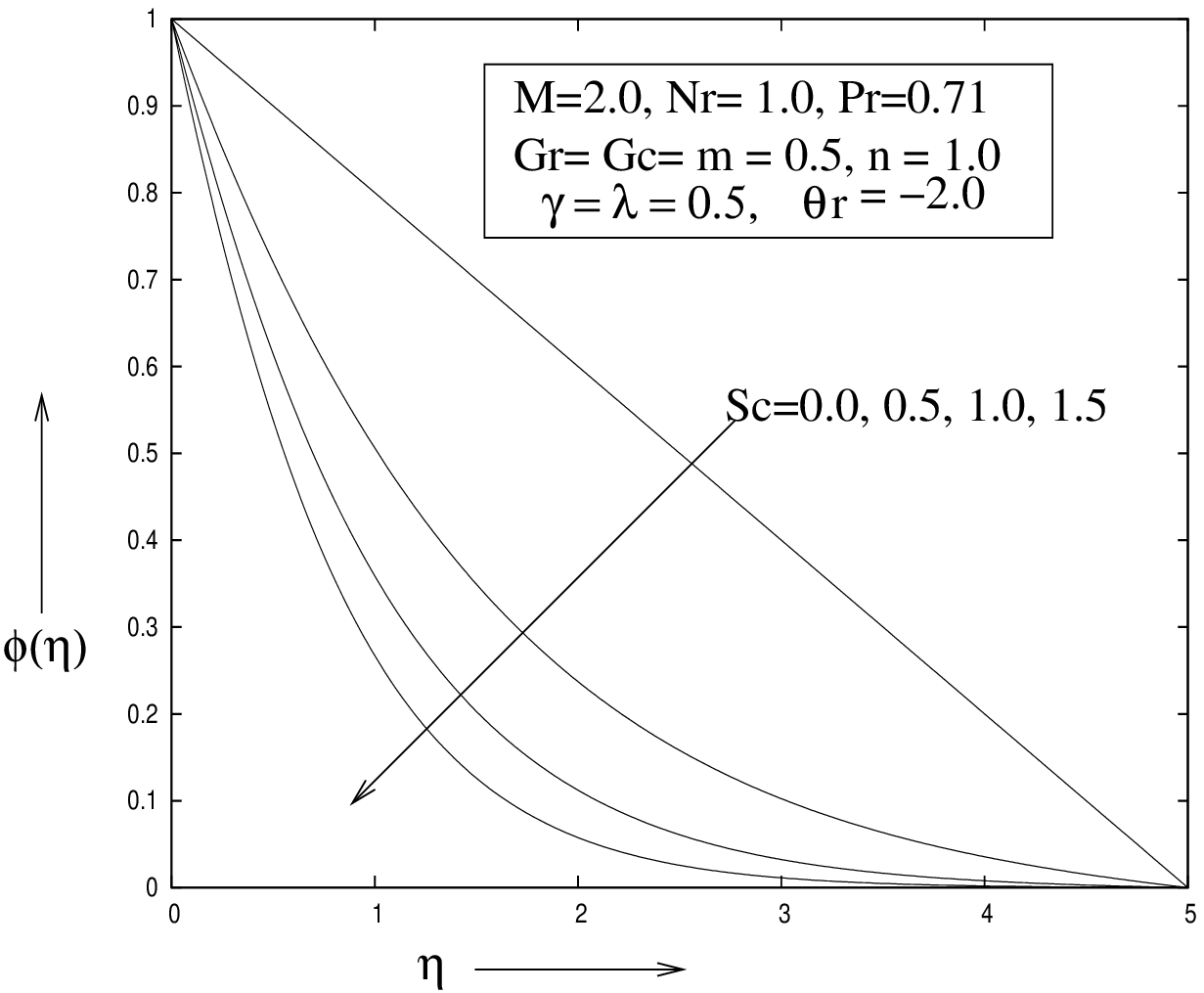}\\
Fig. 29 ~~Influence of concentration species $\phi(\eta)$ with
$\eta$ for different
       values of $Sc$\vspace*{1.5cm} \\

\includegraphics[width=3.8in,height=2.5in ]{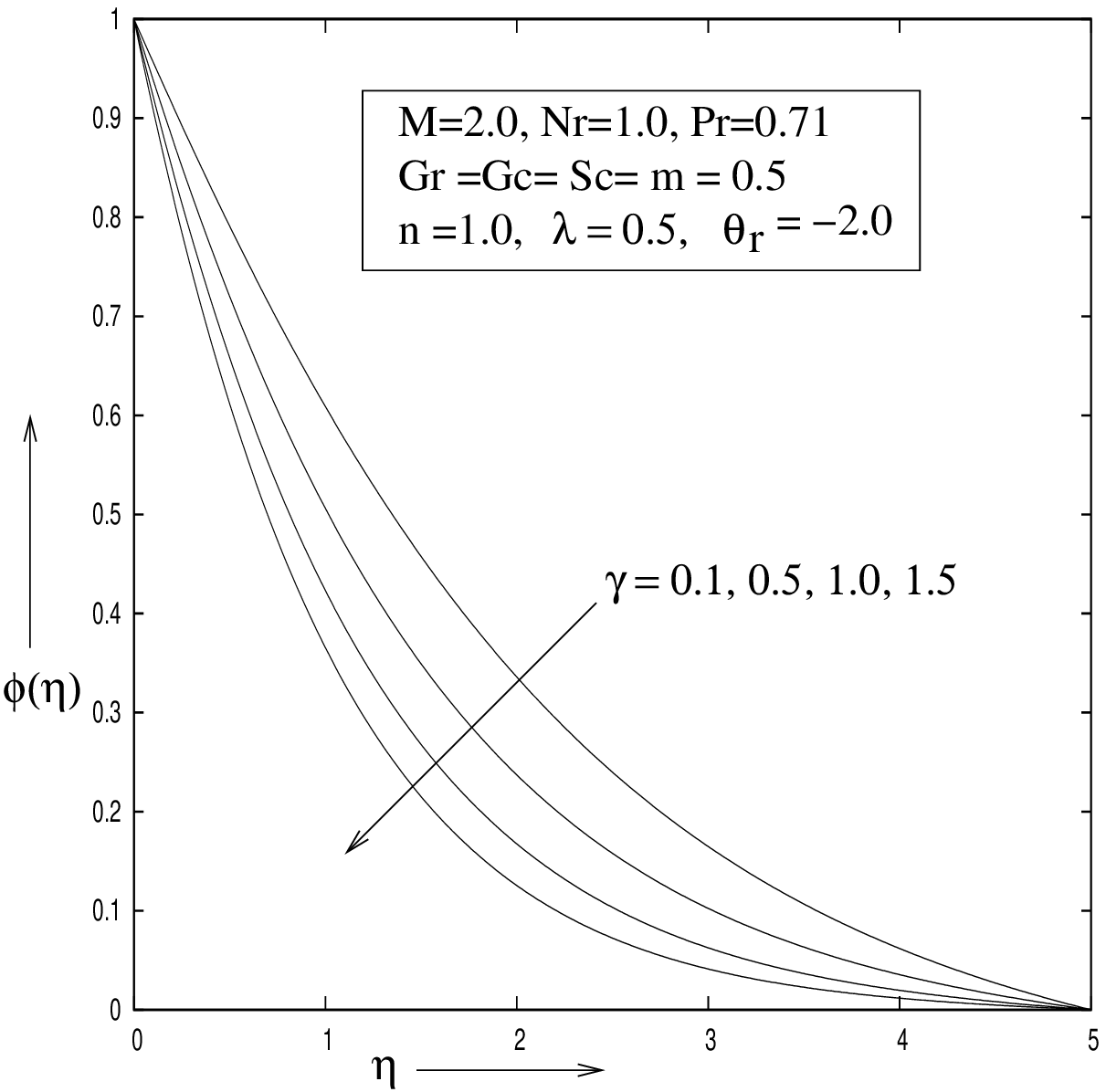}\\
Fig. 30~~ Influence of concentration species $\phi(\eta)$ with
$\eta$ for different
       values of $\gamma$ \\
\end{center}
\end{minipage}\vspace*{.5cm}\\

\begin{minipage}{1.0\textwidth}
   \begin{center}
   \includegraphics[width=3.8in,height=2.5in ]{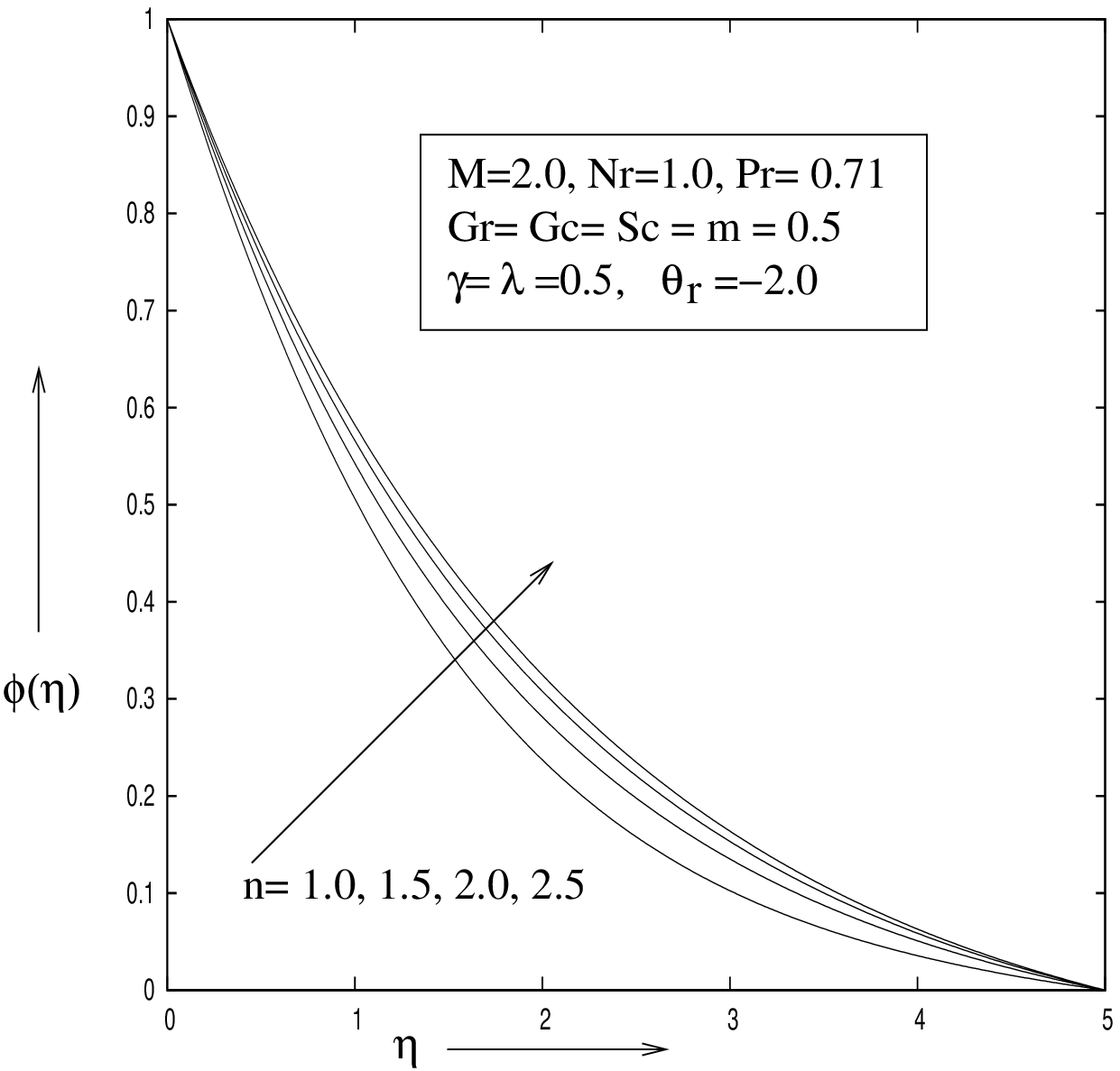}\\
 Fig. 31~~Influence of concentration species $\phi(\eta)$ with
$\eta$ for different
      values of $n$\vspace*{1.5cm} \\
\end{center}
\end{minipage}\vspace*{.5cm}\\

\newpage
Table-1. Numerical values of the local skin-friction coefficient
$C_f$ = $f''(0)$
\begin{center}
\begin{tabular} {|c c c c c c c|c|c|c|}
 \hline
 $Pr~$ & $Nr~ $ & $m~$ & $\theta_r~ $&$\gamma~$&$\lambda~$&$Sc~$ &  $M=0.0$ & $M=2.0$ & $M=4.0$
\\ \hline
    0.71~ & 1.0~ & 0.5~& -2.0~ &0.5~& 0.5~&0.5~&-0.52597&-1.45996&-2.10233 \\
    1.0~  &  & &  & & & & -0.55356& -1.47541 &- 2.11350  \\
    3.0~ &  &  & & & &  & -0.66289 & -1.55043 &-2.17159  \\ \hline
  0.71~ & 0.5 ~& 0.5 ~& -2.0~ &0.5~&0.5~&0.5~ &-0.49627 & -1.44462 &-2.09153 \\
     & 1.0 ~&     &  & &  &  &-0.52596& -1.45996 &- 2.10233  \\
     &5.0 ~&      &  &  &  &  &-0.57792& -1.49009 &- 2.12436 \\
     \hline
  0.71~ & 1.0 ~ & 1.0 ~&-2.0 ~&0.5~&0.5~&0.5~&-0.52596 & -1.22516 & -1.76392 \\
      &     & 3.0~  &  & &  & &- 0.52596 &- 0.73786 & -1.00484  \\
      &     & 5.0 ~& & & & &- 0.52596 & -0.61575 &- 0.75757 \\
 \hline0.71~ &1.0~ & 0.5~ & -1.0~ &0.5~&0.5~&0.5~&-0.56098& -1.66484 &-2.41425 \\
     &   &   & -2.0~   & & & & -0.52596 &- 1.45996 &- 2.10233 \\
      &  &    & -5.0~ & & & &-0.49443 & -1.31365 &- 1.88306\\
 \hline 0.71~& 1.0~&0.5~&-2.0~&0.1~&0.5~&0.5~&-0.49746&-1.44034&-2.08909 \\
     &  &  &   &0.5~  & & &-0.52596&-1.45996&-2.10233 \\
      &   &  &  &1.0~& & &- 0.55088 & -1.47691 &-2.11402 \\
      \hline
0.71~& 1.0~&0.5~&-2.0~&0.5~&-0.5~&0.5~&-0.42172&-1.38284&-2.04079 \\
     &  &   &  & &0.0~& &-0.48498&-1.43029&-.07862 \\
      &  &   & &  &0.5~& &-0.52596 &- 1.45996 &-2.10233 \\
      \hline

0.71~& 1.0~&0.5~&-2.0~&0.5~&-0.5~&0.5~&-0.52596&-1.45996&-2.10233 \\
     &  &   & & & &1.0 &-0.57899&-1.48921&-2.12044 \\
      &  &   & & &  &1.5~&- 0.61068 &- 1.50989 &-2.13432 \\
      \hline

    \end{tabular}
    \end{center}
\newpage
 Table-2. Numerical values of the local Nusselt number $Nu$ =
 $-\theta'(0)$
 \begin{center}
\begin{tabular} {|c c c c c c c|c|c|c|}
 \hline
 $Pr~$ & $Nr~ $ & $m~$ & $\theta_r~ $&$\gamma~$&$\lambda~$&$Sc~$ &  $M=0.0$ & $M=2.0$ & $M=4.0$
\\ \hline
    0.71~ & 1.0~ & 0.5~& -2.0~ &0.5~& 0.5~&0.5~&0.51857&0.48436&0.46751 \\
    1.0~  &  & &  & & & & 0.61651& 0.57475 & 0.55368  \\
    3.0~ &  &  & & & &  & 1.09706 & 1.03107 & 0.99413  \\ \hline
  0.71~ & 0.5 ~& 0.5 ~& -2.0~ &0.5~&0.5~&0.5~ &0.41891 & 0.39371 &0.38159 \\
     & 1.0 ~&     &  & &  &  &0.51857& 0.48436 & 0.46751  \\
     &5.0 ~&      &  &  &  &  &0.70871& 0.66088 & 0.63618 \\
     \hline
  0.71~ & 1.0 ~ & 1.0 ~&-2.0 ~&0.5~&0.5~&0.5~&0.51857 & 0.48973 & 0.47213 \\
      &     & 3.0~  &  & &  & & 0.51857 & 0.50769 & 0.49279  \\
      &     & 5.0 ~& & & & & 0.51857 & 0.51382 & 0.50509 \\
 \hline0.71~ &1.0~ & 0.5~ & -1.0~ &0.5~&0.5~&0.5~&0.51673& 0.48081 &0.46364 \\
     &   &   & -2.0~   & & & & 0.51857 & 0.48436 & 0.46751 \\
      &  &    & -5.0~ & & & & 0.52016 & 0.48729 & 0.47070\\
 \hline 0.71~& 1.0~&0.5~&-2.0~&0.1~&0.5~&0.5~&0.52097&0.48641&0.46901 \\
     &  &  &   &0.5~  & & &0.51857&0.48436&0.46751 \\
      &   &  &  &1.0~& & & 0.51662 & 0.48279 &0.46639 \\
      \hline
0.71~& 1.0~&0.5~&-2.0~&0.5~&-0.5~&0.5~&0.12334&0.02936&-0.02061 \\
     &  &   &  & &0.0~& &0.35575&0.30484&0.27964 \\
      &  &   & &  &0.5~& &0.57857 & 0.48436 &0.46751 \\
      \hline

0.71~& 1.0~&0.5~&-2.0~&0.5~&-0.5~&0.5~&0.51857&0.48436&0.46751 \\
     &  &   & & & &1.0 &0.51425&0.48156&0.46573 \\
      &  &   & & &  &1.5~& 0.51223 & 0.48004 &0.46469 \\
      \hline

    \end{tabular}
 \end{center}
\newpage
 Table-3. Numerical values of the local Sherwood number $Sh$ =
 $-\phi'(0)$
\begin{center}

\begin{tabular} {|c c c c c c c|c|c|c|}
 \hline
 $Pr~$ & $Nr~ $ & $m~$ & $\theta_r~ $&$\gamma~$&$\lambda~$&$Sc~$ &  $M=0.0$ & $M=2.0$ & $M=4.0$
\\ \hline
    0.71~ & 1.0~ & 0.5~& -2.0~ &0.5~& 0.5~&0.5~&0.67199&0.62481&0.60072 \\
    1.0~  &  & &  & & & & 0.66905& 0.62322 & 0.59979  \\
    3.0~ &  &  & & & &  & 0.66032 & 0.61772 & 0.59634  \\ \hline
  0.71~ & 0.5 ~& 0.5 ~& -2.0~ &0.5~&0.5~&0.5~ &0.67544 & 0.62659 &0.60175 \\
     & 1.0 ~&     &  & &  &  &0.67199& 0.62481 & 0.60072  \\
     &5.0 ~&      &  &  &  &  &0.66670& 0.62188 & 0.59898 \\
     \hline
  0.71~ & 1.0 ~ & 1.0 ~&-2.0 ~&0.5~&0.5~&0.5~&0.67199 & 0.63256 & 0.60770 \\
      &     & 3.0~  &  & &  & & 0.67199& 0.65744 & 0.63733  \\
      &     & 5.0 ~& & & & & 0.67199 & 0.66566 & 0.65406 \\
 \hline0.71~ &1.0~ & 0.5~ & -1.0~ &0.5~&0.5~&0.5~&0.66593& 0.61956 &0.59486 \\
     &   &   & -2.0~   & & & & 0.67199 & 0.62481 & 0.60072 \\
      &  &    & -5.0~ & & & & 0.67411 & 0.62911 & 0.60553\\
 \hline 0.71~& 1.0~&0.5~&-2.0~&0.1~&0.5~&0.5~&0.50642&0.43681&0.40050 \\
     &  &  &   &0.5~  & & &0.67199&0.62481&0.60072 \\
      &   &  &  &1.0~& & & 0.83627 & 0.80234 &0.78493 \\
      \hline
0.71~& 1.0~&0.5~&-2.0~&0.5~&-0.5~&0.5~&0.68296&0.63338&0.60670 \\
     &  &   &  & &0.0~& &0.67616&0.62786&0.60279 \\
      &  &   & &  &0.5~& &0.67199 & 0.62481 &0.60072 \\
      \hline

0.71~& 1.0~&0.5~&-2.0~&0.5~&-0.5~&0.5~&0.67199&0.62481&0.60072 \\
     &  &   & & & &1.0 &0.96662&0.90329&0.86889 \\
      &  &   & & &  &1.5~& 1.19569 & 1.12354 &1.08281 \\
      \hline

    \end{tabular}
 \end{center}

\end{document}